\documentclass[%
 aip,
 jcp,%
 amsmath,amssymb,
reprint%
]{revtex4-2}

\usepackage{graphicx}
\usepackage{dcolumn}
\usepackage{bm}
\usepackage{bbm}
\usepackage{hyperref}
\usepackage{comment}
\usepackage{tabularx}
\usepackage{diagbox}
\usepackage{color}

\begin{document}

\title{Numerically ``exact" charge transport dynamics in a dissipative electron--phonon model rationalizing the success of the transient localization scenario}

\author{Veljko Jankovi\'c}
 \affiliation{Institute of Physics Belgrade, University of Belgrade, Pregrevica 118, 11080 Belgrade, Serbia}
 \email{veljko.jankovic@ipb.ac.rs}

\begin{abstract}
Optical conductivity in molecular semiconductors is suppressed in the terahertz region, featuring the displaced Drude peak that reflects carriers' transient localization (TL) by slow intermolecular vibrations.
Meanwhile, recent computations in minimal models evidence optical-conductivity enhancements below the characteristic vibrational frequency, which cannot be captured by the TL phenomenology.
These models assume that the carrier's hopping amplitude is modulated by a single undamped vibration.
The modulation is, however, by many low-frequency modes, whose net effect can be approximated using a few effective damped oscillators.
Here, we employ the dissipaton equations of motion (DEOM) method to compute the finite-temperature real-time current autocorrelation function in a one-dimensional model with Brownian-oscillator spectral density of nonlocal carrier--phonon interaction.
We exploit the dissipaton algebra to handle the phonon-assisted current, reduce the method's computational requirements by working in momentum space, and confirm that numerically stable transport dynamics are virtually independent of a specific DEOM closing scheme.
With increasing damping, we find that DEOM optical-conductivity profiles become increasingly qualitatively similar to TL predictions.
For parameters representative of room-temperature hole transport in single-crystal rubrene, we conclude that the TL phenomenology is established already in the underdamped-oscillator regime.
Reasonable variations in the damping constant weakly affect the carrier mobility, which remains within experimental bounds.
Overall, our results strongly suggest that optical-conductivity enhancements at very low frequencies are artifacts of the assumed delta-like phonon spectrum and rationalize the success of the TL phenomenology in describing experimental data.
\end{abstract}

\maketitle

\section{Introduction}
It is by now well established that the motion of a charge carrier in high-mobility organic semiconductors is mainly limited by its moderate coupling to slow and abundantly thermally excited intermolecular vibrations.~\cite{PhysRevLett.96.086601,PSSB.249.1655,AdvFunctMater.26.2292,PhysStatusSolidiRRL.13.1800485,JChemPhys.152.190902,NatMater.19.491,AngewChemIntEd.64.e202507566}
The carrier's optical response then features a displaced Drude peak (DDP) in the terahertz region,~\cite{ApplPhysLett.89.182103,PhysRevLett.99.016403,ApplPhysLett.105.143302,ApplPhysLett.115.143301,ApplPhysLett.120.053302,NatMater.22.1361} which can be reproduced by the phenomenological Drude--Anderson model~\cite{PhysRevB.89.235201,AdvFunctMater.26.2292} that rests on the TL scenario (TLS).~\cite{PhysRevB.83.081202,PhysRevB.86.245201,AdvFunctMater.26.2292}
Starting from the frozen-phonon limit and introducing effective phonon dynamics in the relaxation-time approximation,~\cite{PhysRevB.83.081202,AdvFunctMater.26.2292} the TLS offers a physically plausible and computationally favorable alternative~\cite{NatMater.16.998,JPhysChemC.123.6989} to quantum--classical simulations of coupled carrier--phonon dynamics.~\cite{PhysRevLett.96.086601,PhysChemChemPhys.17.12395,AccChemRes.55.819,runeson2024}
The latter are considered as the best available approximation to fully quantum dynamics, which are prohibitively expensive in the physically relevant slow-phonon regime.~\cite{JPhysChemA.110.4065,AdvMater.19.2000,JChemPhys.152.190902}
This holds true even in the one-dimensional transport model featuring a single vibrational mode per lattice site.~\cite{PhysRevLett.96.086601,AdvFunctMater.26.2292}

Recent quantum--classical~\cite{runeson2024} as well as our fully quantum numerically ``exact''~\cite{part1,part2} computations in the above-described one-dimensional Peierls (or Su--Schrieffer--Heeger)~\cite{PhysRevLett.42.1698,PhysRevB.56.4484,PhysRevLett.105.266605} model have suggested that TLS predictions are qualitatively inaccurate on long timescales (at low frequencies).
For model parameters appropriate for room-temperature hole transport along the maximum-conductivity direction in single crystals of rubrene,~\cite{AdvMater.19.2000}the authors of Refs.~\onlinecite{part2} and~\onlinecite{runeson2024} evidence optical-response enhancements at very low frequencies.
Meanwhile, most experimental low-frequency signatures align with TLS (Drude--Anderson) predictions,~\cite{PhysRevLett.99.016403,ApplPhysLett.105.143302,ApplPhysLett.115.143301,ApplPhysLett.120.053302} displaying a steady suppression of the optical response below the DDP, though there are exceptions.~\cite{ApplPhysLett.89.182103}

The aforementioned minimal-model studies~\cite{part2,runeson2024} assume that the carrier hopping amplitude is modulated by a single phonon mode (also known as the ``killer" mode~\cite{AngewChemIntEd.64.e202507566,AdvMater.31.201902407}) of frequency $\omega_0\approx 50\:\mathrm{cm}^{-1}$.
Although the ``killer" mode does have the dominant contribution to the total dynamic disorder, a more realistic model should acknowledge the presence of other low-frequency modes, typically quasi-continuously distributed over the $(0-200)\:\mathrm{cm}^{-1}$ range.~\cite{PhysRevLett.102.116602}
A viable strategy to tackle such a model is to assume that only a few (or even a single) effective oscillators are actually coupled to the carrier and to incorporate the effect of the remaining modes by subjecting the effective oscillators' dynamics to a random force and the corresponding (via the fluctuation--dissipation relation) friction.~\cite{PhysRevLett.102.116602,JChemPhys.131.014703}
Indeed, it was shown that the spectrum of the transfer-integral autocorrelation function emerging from molecular-dynamics simulations of a discotic liquid crystal~\cite{JChemPhys.129.094506,PhysRevLett.102.116602} can be accurately represented as a sum of three effective oscillators.
While the most dominant oscillator for the material studied in Ref.~\onlinecite{PhysRevLett.102.116602} exhibits overdamped dynamics, one can argue~\cite{JChemPhys.131.014703,runeson2024} that the most relevant oscillator for molecular crystals performs underdamped oscillations.~\cite{AdvMater.19.2000}
In other words, the spectral density (SD) of the carrier--phonon interaction,~\cite{Mukamel-book,Valkunas-Abramavicius-Mancal-book} which is the only phonon-dependent quantity determining purely carrier dynamics under standard assumptions,~\cite{AnnPhys.24.118} can be reasonably described within the Brownian-oscillator (BO) model.~\cite{JChemPhys.83.4491,PhysRevB.30.1208,Mukamel-book}

When approached from the perspective of open quantum dynamics,~\cite{Breuer-Petruccione-book,Weiss-book} phonon-limited charge transport is usually studied by tracking the relaxation of an initial nonequilibrium charge distribution toward equilibrium.~\cite{JChemPhys.132.081101,JChemPhys.150.234101,JChemPhys.151.044105,JPhysChemLett.15.1382,JChemPhys.161.084118}
However, such approaches cannot access the time-dependent diffusion constant or frequency-dependent mobility,~\cite{ChemSci.15.16715,ProcNatlAcadSci.122.e2424582122} which are related to the finite-temperature time-dependent current autocorrelation function.~\cite{Mahanbook,Kubo-noneq-stat-mech-book}
Methods of open quantum dynamics have recently been used to compute the autocorrelation function of the purely electronic current operator within the Holstein model.~\cite{JChemPhys.142.174103,JChemPhys.156.244102,JChemPhys.161.084118,ChemSci.15.16715,JChemTheoryComput.20.7052,ProcNatlAcadSci.122.e2424582122}
Analogous computations for the Peierls model are difficult because one has to express the expectation values involving the phonon-assisted current~\cite{PhysRev.150.529,MolPhys.18.49,PhysRevB.69.075212,JChemPhys.142.174103} in terms of purely electronic quantities remaining upon integrating phonons out.
Focusing on models with a \emph{finite} number of \emph{undamped} phonon modes,~\cite{PhysRevLett.96.086601,AdvFunctMater.26.2292,runeson2024} which lack explicit dissipation, we have recently succeeded in doing so using the hierarchical equations of motion (HEOM) method.~\cite{part1,part2}
The crux of our solution is the explicit expression of HEOM auxiliaries in terms of phonon creation and annihilation operators.~\cite{part1}
In models with a continuous distribution of phonon frequencies,~\cite{JChemPhys.132.081101,JChemPhys.151.044105,JChemPhys.161.084118} which are referred to as dissipative, such an expression is not yet available, to the best of our knowledge.~\cite{JChemPhys.145.204109,su2025nonperturbativeopenquantumdynamics}

Here, we rely on the DEOM formalism~\cite{JChemPhys.140.054105,FrontPhys.11.110306,MolPhys.116.780,ChemPhys.515.94,JChemPhys.157.170901} to obtain numerically ``exact'' real-time quantum results in a dissipative Peierls model,~\cite{runeson2024,PhysRevLett.102.116602,JChemPhys.131.014703} which one can use to assess the TLS and quantum--classical approximations.~\cite{runeson2024}
While the dynamical equations of the DEOM formalism are identical to those of the HEOM formalism, the former provides the generalized Wick's theorem,~\cite{JChemPhys.140.054105,FrontPhys.11.110306,MolPhys.116.780,ChemPhys.515.94,JChemPhys.157.170901} which we use to handle the phonon-assisted current.
We concentrate on model parameters representative of room-temperature hole transport along the maximum-conductivity direction in single crystals of rubrene,~\cite{AdvMater.19.2000,AdvFunctMater.26.2292}presenting one of the first applications of the generalized Wick's theorem in an extended model system.
In contrast to most open quantum dynamics-based studies,~\cite{JChemPhys.132.081101,JChemPhys.151.044105,JChemPhys.161.084118} which employ the Drude--Lorentz (DL) SD to model the interaction of a carrier with low-frequency phonons,~\cite{Mukamel-book} we assume the more formally appropriate~\cite{JPhysSocJpn.89.015001} BO SD, whose underdamped variant is also more physically plausible for molecular crystals.~\cite{JChemPhys.131.014703}
We circumvent potential long-time numerical instabilities~\cite{JChemPhys.156.064107} by applying an appropriate DEOM closing scheme,~\cite{JChemPhys.159.094113,part1} which enables us to reliably access the diffusive dynamics and thus the low-frequency dynamical mobility.

For weakly damped phonons (a narrow spectrum of relevant modes), subdiffusive carrier dynamics remain limited to intermediate timescales, similarly to our HEOM~\cite{part2} and quantum--classical~\cite{runeson2024} results at zero damping.
The diffusive transport is then approached from the superdiffusive side, giving rise to a dynamical-mobility enhancement below the phonon frequency.
As the damping is increased (the spectrum of relevant modes widens), the diffusive transport sets in from the subdiffusive side, and DEOM results are qualitatively similar to TLS predictions and experimental results.
We find that this change in the character of the long-time dynamics and low-frequency optical response occurs in the region of underdamped BO, at dampings such that the central frequency of the relevant-phonon spectrum remains well defined.
This rationalizes the success of the TLS in reproducing experimental low-frequency dynamical-mobility signatures in organic crystals, in which relevant phonon modes are narrowly distributed around the ``killer'' mode.~\cite{AdvMater.19.2000,JChemPhys.131.014703}
Physically reasonable variations in the damping constant (the width of the relevant-phonon spectrum) weakly affect the dc mobility, which remains within experimental bounds and whose value can be reproduced reasonably well with the usual choice of the free parameter of the TLS.~\cite{PhysRevResearch.2.013001,PhysRevLett.132.266502}
We relate this parameter to the damping parameter of the BO SD and find that the overall quality of TLS predictions for dynamical mobility improves with increasing damping.
Then, the BO SD becomes increasingly similar to the DL SD,~\cite{JChemPhys.136.224103} and the need to treat the phonon-assisted current permits us to expose formal deficiencies of the DL SD,~\cite{JPhysSocJpn.89.015001} which are not apparent in the related Holstein model,~\cite{JChemPhys.150.234101,ChemSci.15.16715,ProcNatlAcadSci.122.e2424582122,JChemPhys.142.174103,JChemPhys.156.244102,JChemTheoryComput.20.7052} and showcase an application of the procedure to circumvent them.~\cite{JPhysSocJpn.89.015001}
Overall, our fully quantum results provide strong evidence that the long-time superdiffusive dynamics and low-frequency dynamical-mobility enhancements reported in minimal models~\cite{part2,runeson2024} are artifacts of the assumed delta-like phonon spectrum (undamped phonon dynamics).

This paper is organized as follows.
Section~\ref{Sec:model-metod} introduces the model and the DEOM formalism.
Section~\ref{Sec:technical-details} discusses some technical aspects necessary to obtain reliable DEOM results, which we present in Sec.~\ref{Sec:results}.
We summarize our main conclusions in Sec.~\ref{Sec:summary-outlook}.

\section{Model and method}
\label{Sec:model-metod}
\subsection{Model}
We study phonon-limited carrier transport in a dissipative version of the one-dimensional model considered in Refs.~\onlinecite{PhysRevLett.96.086601,AdvFunctMater.26.2292,runeson2024,part1,part2}.
The motion of a carrier along an $N$-site chain with periodic boundary conditions is affected by the interaction with a bath of harmonic oscillators, instead of $N$ undamped oscillators considered in Refs.~\onlinecite{PhysRevLett.96.086601,AdvFunctMater.26.2292,runeson2024,part1,part2}.
In the following, we set the elementary charge $e_0$, the lattice constant $a_l$, and physical constants $\hbar$ and $k_B$ to unity.

The total Hamiltonian in momentum space reads
\begin{equation}
\label{Eq:def_H_tot_momentum}
\begin{split}
    H_\mathrm{tot}&=H_S+H_B+H_{S-B}\\&=\sum_k\varepsilon_k\:|k\rangle\langle k|+\sum_{q\xi}\omega_\xi b_{q\xi}^\dagger b_{q\xi}+\sum_q V_qB_q.
\end{split}
\end{equation}
In Eq.~\eqref{Eq:def_H_tot_momentum}, the wavenumbers of the carrier ($k$) and bath oscillators ($q$) can assume $N$ allowed values $\frac{2\pi n}{N}\in(-\pi,\pi]$.
The carrier Hamiltonian $H_S$ describes a free-carrier band with dispersion $\varepsilon_k=-2J\cos k$ originating from nearest-neighbor hops of amplitude $J$.
The bath Hamiltonian $H_B$ describes $N$ identical sets of harmonic oscillators, which are associated with individual sites and counted by index $\xi$.
The interaction Hamiltonian $H_{S-B}$ depends on the carrier operator
\begin{equation}
\label{Eq:def_V_q}
    V_q=\sum_k M(k,q)|k+q\rangle\langle k|,
\end{equation}
with
\begin{equation}
\label{Eq:def_M_k_q}
    M(k,q)=-2i\left[\sin(k+q)-\sin k\right],
\end{equation}
and the bath operator (with $\overline{q}=-q$)
\begin{equation}
\label{Eq:def_B_q}
    B_q=\sum_{\xi}\frac{g_\xi}{\sqrt{N}}\left(b_{q\xi}+b_{\overline{q}\xi}^\dagger\right).
\end{equation}
The carrier--bath matrix element $M(k,q)$ in Eq.~\eqref{Eq:def_M_k_q} describes linear modulation of nearest-neighbor hopping amplitudes by the difference between displacements of the corresponding local oscillators~\cite{JChemPhys.100.2335,PhysRevLett.96.086601,PhysRevLett.105.266605,AdvFunctMater.26.2292} [see also Eq.~\eqref{Eq:def_H_S_B_real_space}].
Equation~\eqref{Eq:def_M_k_q} implies that bath modes with $q=0$ are uncoupled from the rest of the system.
In the following, we understand that the $q=0$ term is omitted from all sums over phonon wave numbers.

\subsection{Spectral density of carrier--phonon interaction}
\label{SSec:spectral_density}
In the model specified by Eq.~\eqref{Eq:def_H_tot_momentum}, the influence of the bath on carrier dynamics is fully captured by the SD of carrier--bath interaction
\begin{equation}
\label{Eq:def_SD}
    \mathcal{J}(\omega)=\pi\sum_{\xi}g_\xi^2[\delta(\omega-\omega_\xi)-\delta(\omega+\omega_\xi)].
\end{equation}
In models without explicit dissipation,~\cite{PhysRevLett.96.086601,AdvFunctMater.26.2292,runeson2024,part1,part2} $\mathcal{J}(\omega)$ consists of delta peaks at $\pm\omega_0$, where $\omega_0$ is the frequency of the undamped vibrational mode.
Here, we mainly consider the BO SD~\cite{Mukamel-book,Valkunas-Abramavicius-Mancal-book,JPhysSocJpn.78.073802,JChemPhys.132.214502}
\begin{equation}
\label{Eq:J_BO_w}
    \mathcal{J}_\mathrm{BO}(\omega)=2E_0\frac{\omega_0^2\cdot 2\gamma_0\omega}{(\omega_0^2-\omega^2)^2+(2\gamma_0\omega)^2}
\end{equation}
which, in the limit of zero damping $\gamma_0$, reduces to the delta-like SD studied in Refs.~\onlinecite{PhysRevLett.96.086601,AdvFunctMater.26.2292,runeson2024,part1,part2}.
The quantity
\begin{equation}
 E_0=\sum_\xi\frac{g_\xi^2}{\omega_\xi}=\int_{-\infty}^{+\infty}\frac{d\omega}{2\pi}\frac{\mathcal{J}_\mathrm{BO}(\omega)}{\omega}
\end{equation}
is one half of the zero-temperature polaron binding energy in the two-site version of this model, which is analytically solvable.~\cite{PhysRevB.56.4484}
We then define the dimensionless carrier--phonon interaction constant as
\begin{equation}
\label{Eq:def_lambda}
 \lambda=\frac{2E_0}{J}.
\end{equation}

\begin{figure}
    \centering
    \includegraphics[width=\columnwidth]{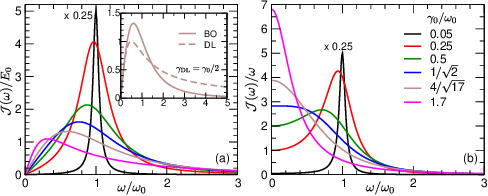}
    \caption{Frequency profile of (a) the BO SD [Eq.~\eqref{Eq:J_BO_w}] and (b) the BO frictional spectrum $\mathcal{J}_\mathrm{BO}(\omega)/\omega$ for different values of the damping $\gamma_0$.
    The curves for $\gamma_0/\omega_0=0.05$ are scaled down by a factor of 4 for visual clarity.
    The inset in (a) compares the BO SD for $\gamma_0/\omega_0=4/\sqrt{17}$ to the DL SD [Eq.~\eqref{Eq:J_DL_omega}] with $\gamma_\mathrm{DL}=\gamma_0/2$.}
    \label{Fig:fig_sd_200525}
\end{figure}

As the SD conveniently combines information on the carrier--phonon interaction strength and phonon density of states, see Eq.~\eqref{Eq:def_SD}, the BO SD can be considered as a convenient model of the influence of a continuous (as opposed to the delta-like~\cite{PhysRevLett.96.086601,AdvFunctMater.26.2292,runeson2024,part1,part2}) density of phonon states on carrier dynamics.
Additionally, the BO model was argued to be relevant for one-dimensional organic semiconductors because it smoothly connects limiting transport regimes in the field of purely dynamic ($\gamma_0/\omega_0\to 0$) and purely static ($\gamma_0/\omega_0\to\infty$) disorder.~\cite{JChemPhys.131.014703,ChemPhysChem.11.2067}
The changes in the BO SD and the so-called frictional spectrum $\mathcal{J}_\mathrm{BO}(\omega)/\omega$~\cite{AnnuRevPhysChem.56.187} with $\gamma_0$ are summarized in Figs.~\ref{Fig:fig_sd_200525}(a) and ~\ref{Fig:fig_sd_200525}(b), respectively.
The frictional spectrum, which essentially determines the free-bath propagator at high temperatures [Eq.~\eqref{Eq:def_mathcalC_t} with $\beta\omega\ll 1$ whenever $\mathcal{J}_\mathrm{BO}(\omega)$ is appreciable],~\cite{JChemPhys.131.014703} peaks at a non-zero (zero) frequency for $\gamma_0<\omega_0/\sqrt{2}$ ($\gamma_0\geq\omega_0/\sqrt{2}$).~\cite{JChemPhys.136.224103}
The underdamped regime $\gamma_0/\omega_0<1$, in which the width of the relevant spectrum (proportional to $\gamma_0$) is such that the vibrational frequency remains well defined, is relevant for organic crystals.~\cite{AdvMater.19.2000,JChemPhys.131.014703}
The overdamped regime $\gamma_0/\omega_0>1$ is relevant for more statically disordered systems.~\cite{JChemPhys.131.014703,PhysRevLett.102.116602}

Many studies aiming at a fundamental understanding of carrier dynamics in organic semiconductors employ the DL SD~\cite{JChemPhys.132.081101,JChemPhys.151.044105,JChemPhys.161.084118}
\begin{equation}
\label{Eq:J_DL_omega}
    \mathcal{J}_\mathrm{DL}(\omega)=2E_0\frac{\omega\gamma_{\mathrm{DL}}}{\omega^2+\gamma_\mathrm{DL}^2},
\end{equation}
where $\gamma_\mathrm{DL}$ is the characteristic bath frequency.
In contrast to the high-frequency behavior $\mathcal{J}_\mathrm{BO}(\omega)\propto\omega^{-3}$ of the BO SD, the DL SD exhibits a long high-frequency tail $\mathcal{J}_\mathrm{DL}(\omega)\propto\omega^{-1}$.
Therefore, even in the high-temperature regime $2\pi T\gg\gamma_\mathrm{DL}$, to which the typically used values of $\gamma_\mathrm{DL}=40\:\mathrm{cm}^{-1}$ and $T=300\:\mathrm{K}=210\:\mathrm{cm}^{-1}$ belong,~\cite{JChemPhys.132.081101,JChemPhys.151.044105,JChemPhys.161.084118} the long tail of the DL SD implies that HEOM/DEOM computations have to take into account a number of Matsubara terms to achieve full convergence.~\cite{JPhysSocJpn.89.015001}
To reduce that number, Ishizaki~\cite{JPhysSocJpn.89.015001} proposed that the DL SD in Eq.~\eqref{Eq:J_DL_omega} be replaced with the BO SD in Eq.~\eqref{Eq:J_BO_w} with $\gamma_0=2\gamma_\mathrm{DL}$ and $\omega_0^2=\gamma_0^2+\epsilon^2$, with $|\epsilon|\lesssim\gamma_0/4$.
The inset of Fig.~\ref{Fig:fig_sd_200525}(a) compares the DL SD to the corresponding BO SD for $|\epsilon|=\gamma_0/4$, so that $\gamma_0/\omega_0=4/\sqrt{17}$.
In addition to lowering the number of Matsubara terms, Ishizaki's approximation to DL SD circumvents unphysical short-time dynamics of mixed carrier--phonon quantities, which we discuss in Sec.~\ref{SSec:transport-DL}.

\subsection{Dissipaton decomposition}
The sole bath quantity influencing the reduced carrier dynamics is the free-bath propagator ($t>0$)~\cite{AnnPhys.24.118}
\begin{equation}
\label{Eq:BCF_q2_q1_infinite_sum}
\begin{split}
    \left\langle B_{q_2}^{(I)}(t)B_{q_1}^{(I)}(0)\right\rangle_B&=\frac{\delta_{q_1\overline{q_2}}}{N}\mathcal{C}(t),
\end{split}
\end{equation}
where $\langle\cdot\rangle_B$ denotes averaging over the equilibrium state $\rho_B^\mathrm{eq}=e^{-\beta H_B}/\mathrm{Tr}_B\:e^{-\beta H_B}$ of the bath at temperature $T=\beta^{-1}$, $B_q^{(I)}(t)=e^{iH_Bt}B_qe^{-iH_Bt}$, while
\begin{equation}
\label{Eq:def_mathcalC_t}
\begin{split}
    \mathcal{C}(t)&=\int_{-\infty}^{+\infty}\frac{d\omega}{\pi}e^{-i\omega t}\frac{\mathcal{J}_\mathrm{BO}(\omega)}{1-e^{-\beta\omega}}\approx\sum_{m=0}^{K-1}c_me^{-\mu_mt}.
\end{split}
\end{equation}
The exponential decomposition of $\mathcal{C}(t)$ involves a total of $K=N_\mathcal{J}+N_\mathrm{BE}$ terms, $N_\mathcal{J}$ ($N_\mathrm{BE}$) of which stem from the poles of the BO SD (the Bose--Einstein factor) in the lower half plane (so that $\mathrm{Re}\:\mu_m>0$).
The coefficients $c_m$ are in general complex, while the coefficients $\mu_m$ appear either purely real or in complex-conjugated pairs,~\cite{MolPhys.116.780} which motivates the definition of index $\overline{m}$ by $\mu_{\overline{m}}=\mu_m^*$.
The values of $c_m$ and $\mu_m$ for the BO SD are summarized in Appendix~\ref{App:c_m_mu_m}.

Equation~\eqref{Eq:def_mathcalC_t} motivates the introduction of dissipaton operators~\cite{MolPhys.116.780,JChemPhys.157.170901} $f_{qm}$ such that $B_q\approx\sum_{m=0}^{K-1}f_{qm}$, while
\begin{equation}
\label{Eq:dissipaton_CF_fwd}
    \left\langle f_{q_2m_2}^{(I)}(t)f_{q_1m_1}^{(I)}(0)\right\rangle_B=\delta_{m_1\overline{m_2}}\eta_{q_2q_1m_2}e^{-\mu_{m_2}t},
\end{equation}
with $\eta_{q_2q_1m}=\delta_{q_1\overline{q_2}}c_m/N$.
Equation~\eqref{Eq:dissipaton_CF_fwd} somewhat differs from the standard DEOM prescription, which features $\delta_{m_1m_2}$ instead of $\delta_{m_1\overline{m_2}}$ [see, e.g., Eq.~(3.2) of Ref.~\onlinecite{MolPhys.116.780} or Eq.~(8) of Ref.~\onlinecite{JChemPhys.157.170901}].
Our choice in Eq.~\eqref{Eq:dissipaton_CF_fwd} is motivated by the requirement that the dissipative theory smoothly reduces to the dissipationless case as $\gamma_0\to 0$.
Then, we have demonstrated~\cite{part1} that $f_{qm}$ are proportional to the creation and annihilation operators $b_q$ and $b_{\overline{q}}^\dagger$ of a single vibrational quantum. 
The same continuity argument implies that $f_{qm}^\dagger=f_{\overline{q}\:\overline{m}}$, while the results of Ref.~\onlinecite{MolPhys.116.780} suggest that $f_{qm}^\dagger=f_{\overline{q}m}$.
We emphasize that these subtle differences between our and original DEOM formulations exist only in the underdamped regime,~\cite{JChemPhys.145.204109} in which $\mu_m$s appear in complex-conjugated pairs.
While we expect that the differences would affect dissipaton-resolved quantities, the quantities depending on $f_{qm}$ only through their sum $B_q$ [see, e.g., Eq.~\eqref{Eq:def_j_e-ph}] do not depend on whether Eq.~\eqref{Eq:dissipaton_CF_fwd} features $\delta_{m_1\overline{m_2}}$ or $\delta_{m_1m_2}$.
Most of the previous DEOM-based studies dealt with the (strongly) overdamped regime,~\cite{JChemPhys.142.024112,MolPhys.116.780,ChemPhys.515.94,JChemPhys.154.244105} in which all $\mu_m$s are real ($\overline{m}=m$), so that Eq.~\eqref{Eq:dissipaton_CF_fwd} reduces to the standard DEOM prescription.
The DEOM construction also needs the time-reversal counterpart of Eq.~\eqref{Eq:dissipaton_CF_fwd}, which reads
\begin{equation}
\label{Eq:dissipaton_CF_bwd}
\left\langle f_{q_1m_1}^{(I)}(0)f_{q_2m_2}^{(I)}(t)\right\rangle_B=\delta_{m_1\overline{m_2}}\eta_{\overline{q_2}\:\overline{q_1}\:\overline{m_2}}^*e^{-\mu_{m_2}t}.
\end{equation}

\subsection{DEOM-based framework for transport properties}
Carrier transport dynamics are encoded in the finite-temperature real-time current--current correlation function~\cite{Mahanbook,Kubo-noneq-stat-mech-book}
\begin{equation}
\label{Eq:def_C_jj}
    C_{jj}(t)=\mathrm{Tr}\left\{je^{-iH_\mathrm{tot}t}j\rho_\mathrm{tot}^\mathrm{eq}e^{iH_\mathrm{tot}t}\right\},
\end{equation}
where $\rho_\mathrm{tot}^\mathrm{eq}=e^{-\beta H_\mathrm{tot}}/\mathrm{Tr}\{e^{-\beta H_\mathrm{tot}}\}$.
The current operator $j=j_\mathrm{e}+j_\mathrm{e-ph}$ is the sum of the purely electronic (band) current
\begin{equation}
\label{Eq:def_j_e}
    j_\mathrm{e}=\sum_k v_kP_k,
\end{equation}
with $v_k=\partial_k\varepsilon_k$ and $P_k=|k\rangle\langle k|$,
and the phonon-assisted current
\begin{equation}
\label{Eq:def_j_e-ph}
    j_\mathrm{e-ph}=\sum_q J_qB_q\approx\sum_{kq}\sum_{m=0}^{K-1}M_J(k,q)|k+q\rangle\langle k|f_{qm},
\end{equation}
with $M_J(k,q)=\partial_k M(k,q)$.~\cite{part1,part2}

Within the DEOM theory, the operator
\begin{equation}
\label{Eq:def_iota_tot}
    \iota_\mathrm{tot}(t)=e^{-iH_\mathrm{tot}t}j\rho_\mathrm{tot}^\mathrm{eq}e^{iH_\mathrm{tot}t}
\end{equation}
is represented by the so-called dissipaton density operators (DDOs)~\cite{MolPhys.116.780,JChemPhys.157.170901}
\begin{equation}
\label{Eq:iota_tot_DEOM_represent}
   \iota_\mathbf{n}^{(n)}(t)=\mathrm{Tr}_B\left\{F_\mathbf{n}^{(n)}\iota_\mathrm{tot}(t)\right\}. 
\end{equation}
These describe many-dissipaton configurations labeled by the $(N-1)K$-dimensional vector $\mathbf{n}=[n_{qm}]$ of nonnegative integers $n_{qm}$ such that $\sum_{qm}n_{qm}=n$.
The bath operator $F_\mathbf{n}^{(n)}$ is the irreducible product of single-dissipaton operators $f_{qm}$, which remains invariant under permutations of $f$ operators.~\cite{JChemPhys.157.170901}
We will need
\begin{equation}
\label{Eq:0-1-dissipaton-configs}
    F_\mathbf{0}^{(0)}=\mathbb{I}_B,\quad F_{\mathbf{0}_{qm}^+}^{(1)}=f_{qm}.
\end{equation}
Although we have succeeded in expressing $F_\mathbf{n}^{(n)}$ for $n\geq 2$ in terms of $f_{qm}$ in the undamped case,~\cite{part1} we note that this task is considerably more difficult and still incompletely solved in the dissipative case considered here.~\cite{JChemPhys.145.204109,su2025nonperturbativeopenquantumdynamics}

The operators $\iota_\mathbf{n}^{(n)}(t)$ satisfy the real-time DEOM (r-DEOM)~\cite{MolPhys.116.780,JChemPhys.157.170901}
\begin{equation}
\label{Eq:r-DEOM_before_rescaling}
\begin{split}
    &\partial_t\iota_\mathbf{n}^{(n)}(t)=-i(H_S^\times-i\mu_\mathbf{n})\iota_\mathbf{n}^{(n)}(t)-i\sum_{qm}V_q^\times \iota_{\mathbf{n}_{qm}^+}^{(n+1)}(t)\\
    &-i\sum_{qm}n_{qm}\sum_{q'}\left(\eta_{qq'm}V_{q'}^>-\eta_{\overline{q}\:\overline{q'}\overline{m}}^*V_{q'}^<\right)\iota_{\mathbf{n}_{qm}^-}^{(n-1)}(t).
\end{split}
\end{equation}
In Eq.~\eqref{Eq:r-DEOM_before_rescaling}, $V^\times O=[V,O]$, $V^>O=VO$, $V^<O=OV$, while $\mu_\mathbf{n}=\sum_{qm}n_{qm}\mu_m$.
At each instant $t$, we combine Eqs.~\eqref{Eq:def_C_jj}--\eqref{Eq:0-1-dissipaton-configs} to obtain
\begin{equation}
\begin{split}
    &C_{jj}(t)=\mathrm{Tr}\{j\iota_\mathrm{tot}(t)\}=\sum_k v_k\langle k|\iota_\mathbf{0}^{(0)}(t)|k\rangle\\&+\sum_{kqm}M_J(k,q)\langle k|\iota_{\mathbf{0}_{qm}^+}^{(1)}(t)|k+q\rangle.
\end{split}
\end{equation}

The r-DEOM in Eq.~\eqref{Eq:r-DEOM_before_rescaling} is propagated starting from the initial condition
\begin{equation}
\label{Eq:iota_n_init_cond}
    \iota_\mathbf{n}^{(n)}(0)\equiv\iota_\mathbf{n}^{(n,\mathrm{eq})}=\mathrm{Tr}_B\left\{F_\mathbf{n}^{(n)}j\rho_\mathrm{tot}^\mathrm{eq}\right\},
\end{equation}
which is to be expressed in terms of the DDOs $\rho_\mathbf{n}^{(n,\mathrm{eq})}=\mathrm{Tr}_B\{F_\mathbf{n}^{(n)}\rho_\mathrm{tot}^\mathrm{eq}\}$ constituting the DEOM representation of $\rho_\mathrm{tot}^\mathrm{eq}$.
The contribution of $j_\mathrm{e}$ to Eq.~\eqref{Eq:iota_n_init_cond} is
\begin{equation}
    \iota_{\mathbf{n},\mathrm{e}}^{(n,\mathrm{eq})}=\mathrm{Tr}_B\left\{F_\mathbf{n}^{(n)}j_\mathrm{e}\rho_\mathrm{tot}^\mathrm{eq}\right\}=\sum_k v_kP_k\rho_\mathbf{n}^{(n,\mathrm{eq})}.
\end{equation}
To evaluate the contribution of $j_\mathrm{e-ph}$ to Eq.~\eqref{Eq:iota_n_init_cond}, we use the generalized Wick's theorem~\cite{MolPhys.116.780,JChemPhys.157.170901}
\begin{equation}
\label{Eq:GWT_gtr}
\begin{split}
    &\mathrm{Tr}_B\left\{F_\mathbf{n}^{(n)}f_{qm}\rho_\mathrm{tot}^\mathrm{eq}\right\}=\\&\rho_{\mathbf{n}_{qm}^+}^{(n+1,\mathrm{eq})}+\sum_{q'm'}n_{q'm'}\langle f_{q'm'}f_{qm}\rangle_B^>\rho_{\mathbf{n}_{q'm'}^-}^{(n-1,\mathrm{eq})},
\end{split}
\end{equation}
where [see Eq.~\eqref{Eq:dissipaton_CF_fwd}]
\begin{equation}
\label{Eq:def_equal_time_gtr}
 \left\langle f_{q'm'}f_{qm}\right\rangle_B^>=\left\langle f_{q'm'}^{(I)}(0_+)f_{qm}^{(I)}(0)\right\rangle_B=\delta_{m\overline{m'}}\eta_{q'qm'}.   
\end{equation}
Combining Eqs.~\eqref{Eq:def_j_e-ph},~\eqref{Eq:GWT_gtr}, and~\eqref{Eq:def_equal_time_gtr}, we obtain that the contribution of $j_\mathrm{e-ph}$ to Eq.~\eqref{Eq:iota_n_init_cond} reads
\begin{equation}
\begin{split}
    &\iota_{\mathbf{n},\mathrm{e-ph}}^{(n,\mathrm{eq})}=\mathrm{Tr}_B\left\{F_\mathbf{n}^{(n)}j_\mathrm{e-ph}\rho_\mathrm{tot}^\mathrm{eq}\right\}=\\&\sum_{qm}J_q\rho_{\mathbf{n}_{qm}^+}^{(n+1,\mathrm{eq})}+\sum_{qm}n_{qm}\sum_{q'}\eta_{qq'm}J_{q'}\rho_{\mathbf{n}_{qm}^-}^{(n-1,\mathrm{eq})}.
\end{split}
\end{equation}

\subsection{DEOM representation of the correlated carrier--bath equilibrium}
The DEOM representation $\{\rho_\mathbf{n}^{(n,\mathrm{eq})}\}$ of $\rho_\mathrm{tot}^\mathrm{eq}$ follows from the imaginary-time DEOM (i-DEOM)~\cite{JChemPhys.153.154111,JChemPhys.157.170901}
\begin{equation}
\label{Eq:im-time-deom}
\begin{split}
    &\partial_\tau\sigma_\mathbf{n}^{(n)}(\tau)=-(H_S^\times-i\mu_\mathbf{n})\sigma_\mathbf{n}^{(n)}(\tau)\\&-\sum_{qm}V_q^>\sigma_{\mathbf{n}_{qm}^+}^{(n+1)}(\tau)\\&-\sum_{qm}n_{qm}\sum_{q'}\eta_{qq'm}V_{q'}^>\sigma_{\mathbf{n}_{qm}^-}^{(n-1)}(\tau).
\end{split}
\end{equation}
Equation~\eqref{Eq:im-time-deom} is propagated from $\tau=0$ to $\beta$ with the initial condition
\begin{equation}
    \sigma_\mathbf{n}^{(n)}(0)=\delta_{n,0}\delta_{\mathbf{n},\mathbf{0}}\frac{e^{-\beta H_S}}{\mathrm{Tr}_S\:e^{-\beta H_S}},
\end{equation}
after which $\rho_\mathbf{n}^{(n,\mathrm{eq})}$ is obtained as
\begin{equation}
    \rho_\mathbf{n}^{(n,\mathrm{eq})}=\frac{\sigma_\mathbf{n}^{(n)}(\beta)}{\mathrm{Tr}_S\:\sigma_\mathbf{0}^{(0)}(\beta)}.
\end{equation}

\subsection{Model parameters}
\label{SSec:model_parameters}
We focus on the slow-phonon ($\omega_0/J=0.044$), high-temperature ($T/J=0.175$ or $T/\omega_0\approx 4$), intermediate-interaction ($\lambda=0.336$) regime, which is relevant for anisotropic carrier transport in organic molecular crystals.~\cite{AdvMater.19.2000,AdvFunctMater.26.2292}
The damping parameter $\gamma_0$ in Eq.~\eqref{Eq:J_BO_w} can be estimated from combined molecular-dynamics and quantum-chemistry analyses of the vibrational motions that most strongly modulate carrier transport along the maximum-conductivity direction of rubrene~\cite{AdvMater.19.2000} and a discotic liquid crystal.~\cite{PhysRevLett.102.116602}
The net effect of these motions can be conveniently approximated using few BO modes, the most prominent of which can be either underdamped [$\gamma_0/\omega_0\sim 0.25$ in Fig.~1(b) of Ref.~\onlinecite{AdvMater.19.2000}] or overdamped ($\gamma_0/\omega_0=1.7$ in Ref.~\onlinecite{PhysRevLett.102.116602}).
Assuming undamped vibrations, we have recently examined carrier transport for different adiabaticity ratios $\omega_0/J$, temperatures, and interactions.~\cite{part1,part2} 
We do not expect that the damping of vibrational motions will change the trends observed upon varying these parameters.
Therefore, here we mostly concentrate on understanding the effect of damping on transport dynamics, which has not been widely explored so far.

\subsection{Quantities describing carrier transport}
The physical quantities that we use to characterize transport dynamics are the time-dependent diffusion constant 
\begin{equation}
    \mathcal{D}(t)=\frac{1}{2}\frac{d}{dt}\Delta x^2(t)=\int_0^t ds\:\mathrm{Re}\:C_{jj}(s),
\end{equation}
which determines the growth rate of the carrier's mean-square displacement $\Delta x^2(t)=\langle[x(t)-x(0)]^2\rangle$, and the diffusion exponent
\begin{equation}
    \alpha(t)=\frac{2t\mathcal{D}(t)}{\Delta x^2(t)},
\end{equation}
which determines the instantaneous power-law growth of $\Delta x^2(t)$ with time, $\Delta x^2(t)\propto t^{\alpha(t)}$.
We are mostly interested in how the details of the crossover between short-time ballistic transport, for which $\alpha(t)\approx2$, and long-time diffusive transport, for which $\alpha(t)$ approaches unity, change upon varying the damping rate $\gamma_0$.
This ballistic-to-diffusive crossover exhibits periods of superdiffusive [$\alpha(t)>1$] and subdiffusive [$\alpha(t)<1$] carrier dynamics, and we pay special attention to whether its long-time completion is reached from the super- or subdiffusive side.
In experiments, crossover dynamics are usually inferred from the dynamical-mobility profile
\begin{equation}
\label{Eq:def_re_mu_w}
    \mathrm{Re}\:\mu(\omega)=\frac{1-e^{-\beta\omega}}{2\omega}\int_{-\infty}^{+\infty}dt\:e^{i\omega t}C_{jj}(t).
\end{equation}
The dc mobility $\mu_\mathrm{dc}=\mathrm{Re}\:\mu(\omega=0)$ is connected to the long-time limit $\mathcal{D}_\infty$ of $\mathcal{D}(t)$ through the Einstein relation $\mu_\mathrm{dc}=\frac{\mathcal{D}_\infty}{T}$.

\section{Technical details}
\label{Sec:technical-details}
The parameter regime described in Sec.~\ref{SSec:model_parameters} is challenging because the DEOM [Eqs.~\eqref{Eq:im-time-deom} and~\eqref{Eq:r-DEOM_before_rescaling}] have to be solved on long chains, as implied by the small adiabaticity ratio, and truncated at a moderate maximum depth $D$ dictated by the combination of high temperature and intermediate interaction.
Up to now, the (bosonic) dissipaton algebra has been used to study hybrid system--bath dynamics only in few-level systems (monomers and dimers).~\cite{JChemPhys.142.024112,JChemPhys.145.204109,ChemPhys.515.94,MolPhys.116.780,JChemPhys.154.244105}
Meanwhile, the applications of the (bosonic) DEOM formalism to systems with a somewhat larger number of electronic levels focused on purely electronic dynamics.~\cite{JPhysChemB.124.2354,JPhysChemB.125.11884,JChemPhys.157.084119,ACSOmega.9.51228}

To the best of our knowledge, our study is among the first studies exploiting the (bosonic) dissipaton algebra to compute a finite-temperature real-time correlation function of a mixed system--bath operator in a many-level system.
Our computations are made possible by formulating the DEOM in momentum space,~\cite{JChemPhys.159.094113,part1} in which the DDO $\iota_\mathbf{n}^{(n)}(t)$ in Eq.~\eqref{Eq:r-DEOM_before_rescaling} has only $N$ (instead of $N^2$) nonzero matrix elements $\langle k|\iota_\mathbf{n}^{(n)}(t)|k+k_\mathbf{n}\rangle$ and similarly for $\langle k|\sigma_\mathbf{n}^{(n)}(\tau)|k+k_\mathbf{n}\rangle$ in Eq.~\eqref{Eq:im-time-deom}.
In Secs.~\ref{SSec:DEOM_closing}--\ref{SSec:finite_N_D}, we discuss further technical details needed to gain confidence in our DEOM results.

\subsection{DEOM closing scheme}
\label{SSec:DEOM_closing}

\begin{figure*}[htbp!]
    \centering
    \includegraphics[width=\textwidth]{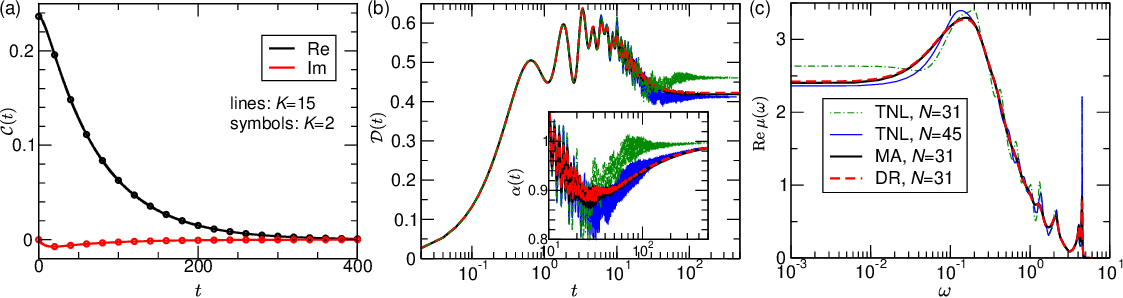}
    \caption{(a) Real (black) and imaginary (red) parts of the bath correlation function $\mathcal{C}(t)$ computed using $K=2$ (symbols) and $K=15$ (lines) terms in the exponential decomposition in Eq.~\eqref{Eq:def_mathcalC_t}.
    The time-dependent diffusion constant (b) and the dynamical-mobility profile (c) computed using DEOM with $\Gamma(k,\mathbf{n})=0$ in Eq.~\eqref{Eq:general_closing} (label ``TNL," thin lines) and with the Markovian-adiabatic [Eqs.~\eqref{Eq:MA_closing} and~\eqref{Eq:def_gamma_k}, label ``MA," thick solid line] and derivative-resum [Eq.~\eqref{Eq:DR_closing}, label ``DR," dashed line] closing terms with respect to $D$.
    The inset of panel (b) compares the diffusion-exponent dynamics computed using the same closing schemes.
    The model parameters are $J=1,\omega_0=0.044,\lambda=0.336,T=0.175,$ and $\gamma_0=1.7\omega_0$, while the maximum DEOM depth is $D=4$.
    }
    \label{Fig:closing}
\end{figure*}

Ideally, the r-DEOM in Eq.~\eqref{Eq:r-DEOM_before_rescaling} is to be solved using $K\to\infty$ terms in the exponential decomposition in Eq.~\eqref{Eq:def_mathcalC_t}, and considering DDOs up to the maximum depth $D\to\infty$.    
In practice, both $K$ and $D$ are finite, and the convergence with respect to them has to be checked.
The convergence can be enhanced by devising appropriate DEOM closing schemes with respect to $K$ and $D$.~\cite{JChemPhys.142.104112,ChinJChemPhys.28.409,MolPhys.116.780}

As we focus on high temperatures, at which $\beta\omega_0\ll 1$ and $\beta\gamma_0\ll 1$ (see Sec.~\ref{SSec:model_parameters}), it is justified to set $K=2$, i.e., to retain only the terms originating from the poles of $\mathcal{J}_\mathrm{BO}(\omega)$ in Eq.~\eqref{Eq:def_mathcalC_t} [see Fig.~\ref{Fig:closing}(a) and Appendix~\ref{App:c_m_mu_m}].
We then employ the Ishizaki--Tanimura closing with respect to $K$,~\cite{JPhysSocJpn.74.3131,JChemPhys.130.084105} augmenting the RHS of Eq.~\eqref{Eq:r-DEOM_before_rescaling} by
\begin{equation}
    [\partial_t\iota_\mathbf{n}^{(n)}(t)]_K=-\Delta\sum_{q} V_q^\times V_{\overline{q}}^\times\iota_\mathbf{n}^{(n)}(t),
\end{equation}
where~\cite{JChemPhys.130.084105}
\begin{equation}
\label{Eq:def_delta}
    \Delta=\sum_{m=K}^{+\infty}\frac{c_m}{\mu_m}=E_0\left(4\frac{T\gamma_0}{\omega_0^2}-i\right)-\sum_{m=0}^{K-1}\frac{c_m}{\mu_m}.
\end{equation}

It is known that the HEOM/DEOM with BO SD can exhibit long-time numerical instabilities,~\cite{JChemPhys.156.064107} similar to those observed in models with undamped vibrations,~\cite{JChemPhys.150.184109,JChemPhys.159.094113} which cannot be removed by simply increasing $D$.
Assuming undamped phonons, for which such instabilities are expected to be the most pronounced, we have shown~\cite{JChemPhys.159.094113,part1,part2} that HEOM can be stabilized by introducing the closing term 
\begin{equation}
\label{Eq:general_closing}
\begin{split}
    &[\partial_t\langle k|\iota_\mathbf{n}^{(n)}(t)|k+k_\mathbf{n}\rangle]_D=\\&-\delta_{n,D}\Gamma(k,\mathbf{n})\langle k|\iota_\mathbf{n}^{(n)}(t)|k+k_\mathbf{n}\rangle
\end{split}
\end{equation}
to the equations at depth $D$.
In contrast to existing approaches,~\cite{JChemPhys.142.104112,ChinJChemPhys.28.409,MolPhys.116.780} which were gauged on few-level systems, the closing term in Eq.~\eqref{Eq:general_closing} does not introduce couplings among either DDOs at depth $D$ or different matrix elements of a DDO at depth $D$.
This feature of Eq.~\eqref{Eq:general_closing} guarantees that the computational cost of the closing remains much smaller than that of one step in DEOM propagation. 
The complex quantity $\Gamma(k,\mathbf{n})$ depends on the details of the closing, i.e., on the approximations employed to solve the equations at depth $D+1$.
Neglecting the couplings to DDOs at depth $D+2$ and using the Markovian and adiabatic approximations yields
\begin{equation}
\label{Eq:MA_closing}
    \Gamma_\mathrm{MA}(k,\mathbf{n})=\frac{1}{2}(\gamma_k+\gamma_{k+k_\mathbf{n}}^*),
\end{equation}
where
\begin{equation}
\label{Eq:def_gamma_k}
\begin{split}
    \gamma_k&=\frac{2}{N}\sum_{qm}\frac{c_m|M(k,q)|^2}{\mu_m+i(\varepsilon_{k+q}-\varepsilon_k)}.
\end{split}
\end{equation}
Another possibility for closing with respect to $D$ is the so-called derivative-resum scheme,~\cite{JChemPhys.142.104112,JChemPhys.157.054108,MolPhys.116.780} which employs
\begin{equation}
\label{Eq:DR_closing}
\begin{split}
    \Gamma_\mathrm{DR}(k,\mathbf{n})&=\frac{1}{N}\sum_{qm}\frac{c_m|M(k,q)|^2}{\mu_\mathbf{n}+\mu_m+i(\varepsilon_{k+q}-\varepsilon_{k+k_\mathbf{n}})}\\
    &+\frac{1}{N}\sum_{qm}\frac{c_{\overline{m}}^*|M(k+k_\mathbf{n},q)|^2}{\mu_\mathbf{n}+\mu_m+i(\varepsilon_k-\varepsilon_{k+k_\mathbf{n}+q})}.
\end{split}
\end{equation}
We have thoroughly presented the procedure to obtain Eqs.~\eqref{Eq:MA_closing}--\eqref{Eq:DR_closing} in Appendix~D of Ref.~\onlinecite{part1}.

Following the practice common in models without explicit dissipation, in Eq.~\eqref{Eq:general_closing}, we retain only the real parts of $\Gamma(k,\mathbf{n})$, i.e., we neglect the closing-induced renormalization of free-oscillation frequencies of DDOs at the terminal layer.
We have checked that the imaginary parts of $\Gamma(k,\mathbf{n})$ in Eq.~\eqref{Eq:general_closing} very weakly affect the overall transport dynamics and, in particular, the long-time transport coefficient $\mu_\mathrm{dc}$.
In undamped models,~\cite{JChemPhys.159.094113,part1} $\mathrm{Re}\:\Gamma_\mathrm{MA}(k,\mathbf{n})$ and $\mathrm{Re}\:\Gamma_\mathrm{DR}(k,\mathbf{n})$ are meaningful only in the infinite-chain limit $N\to\infty$, in which these can be evaluated analytically.
For $\gamma_0\neq 0$, $\mathrm{Re}\:\Gamma_\mathrm{MA}(k,\mathbf{n})$ and $\mathrm{Re}\:\Gamma_\mathrm{DR}(k,\mathbf{n})$ can be computed for any finite $N$, on which they strongly depend.
Although analytical expressions for $\mathrm{Re}\:\Gamma_\mathrm{MA}(k,\mathbf{n})$ and $\mathrm{Re}\:\Gamma_\mathrm{DR}(k,\mathbf{n})$ in the $N\to\infty$ limit are not available for $\gamma_0\neq 0$, we find that their long-chain limits are reached for $N$ much larger than typical chain lengths $N\sim 30$ we consider here.
In practice, we find that computing Eqs.~\eqref{Eq:def_gamma_k} and~\eqref{Eq:DR_closing} on a chain containing $N_\infty=(10^3-10^4)N$ sites provides us with the desired long-chain limits (although $N_\infty/N$ generally decreases with increasing $\gamma_0$).

In Ref.~\onlinecite{part2}, we have checked that both the Markovian-adiabatic [Eqs.~\eqref{Eq:MA_closing} and~\eqref{Eq:def_gamma_k}] and derivative-resum [Eq.~\eqref{Eq:DR_closing}] schemes yield virtually identical transport dynamics for $\gamma_0=0$.
Figures~\ref{Fig:closing}(b) and~\ref{Fig:closing}(c) confirm that the same holds in the overdamped regime $\gamma_0/\omega_0=1.7$.
These also reveal that for sufficiently strong damping, the so-called time-nonlocal closing,~\cite{JChemTheoryComput.8.2808} which sets $\Gamma(k,\mathbf{n})=0$ in Eq.~\eqref{Eq:general_closing}, also yields stable long-time diffusive dynamics from which low-frequency dynamical mobility can be meaningfully extracted.
This stands in sharp contrast to the undamped case, in which computations of long-time transport coefficients crucially require a closing term~\cite{part2} because these cannot be accessed by extrapolating the results for small nonzero dampings to the zero-damping limit.~\cite{JChemPhys.160.111102} 
Although Figs.~\ref{Fig:closing}(b) and~\ref{Fig:closing}(c) suggest that the closing terms can be omitted if $\gamma_0$ is sufficiently large, all our results rely on the propagation of the closed DEOM.
Using one and the same closing scheme throughout the physically reasonable range for $\gamma_0$ down to $\gamma_0=0$ enables a consistent and fair assessment of trends in transport dynamics with varying $\gamma_0$.

Furthermore, Figs.~\ref{Fig:closing}(b) and~\ref{Fig:closing}(c) suggest that our DEOM closing enhances not only the convergence with respect to the maximum depth but also with respect to the chain length.
In particular, the DEOM dynamics with $N=31$ and Markovian-adiabatic or derivative-resum closing are very similar to those employing a $\approx 50\%$-longer chain ($N=45$) and time-nonlocal closing.
In the overdamped regime considered here, the diffusion-constant growth on intermediate timescales and the corresponding dynamical-mobility feature at $\omega\approx\omega_0$ reflect finite-size effects [see Figs.~\ref{Fig:closing}(b) and~\ref{Fig:closing}(c) for $N=31$].
Meanwhile, analogous features for weak damping are not artifacts due to insufficient chain length,~\cite{part2} as discussed in greater detail in Sec.~\ref{SSec:finite_N_D}.  

Finally, we prefer the Markovian-adiabatic [Eqs.~\eqref{Eq:MA_closing} and~\eqref{Eq:def_gamma_k}] to the derivative-resum [Eq.~\eqref{Eq:DR_closing}] scheme because of its more favorable computational requirements. 
In particular, the Markovian-adiabatic scheme needs only $N$ quantities $\gamma_k$ computed on an $N_\infty$-site chain, while the derivative-resum scheme needs a much larger number of quantities $\Gamma_\mathrm{DR}(k,\mathbf{n})$, $N$ quantities for each DDO at depth $D$.
Although all these quantities can be computed only once, before the r-DEOM propagation, our derivative-resum scheme requires much more computational time and computer memory to compute and store the closing terms than the Markovian-adiabatic scheme.
It is for this reason that we employ the Markovian-adiabatic scheme in the following.

\subsection{Thermodynamic quantities}
\label{SSec:thermodynamics_OSR}
The i-DEOM in Eq.~\eqref{Eq:im-time-deom} is constructed by the analytical continuation of the dissipaton algebra,~\cite{JChemPhys.153.154111} which amounts to setting $t=-i\tau$ ($0\leq\tau\leq\beta$) in Eq.~\eqref{Eq:def_mathcalC_t}.
The first equality in Eq.~\eqref{Eq:def_mathcalC_t} shows that the function $\widetilde{\mathcal{C}}(\tau)=\mathcal{C}(-i\tau)$ is purely real and symmetric around $\beta/2$, i.e., $\widetilde{\mathcal{C}}(\beta-\tau)=\widetilde{\mathcal{C}}(\tau)$.
However, the exponential decomposition in Eq.~\eqref{Eq:def_mathcalC_t} does not satisfy these two requirements for finite values of $K$.
For $K=2$, on which we focus here, the exponential decomposition satisfies
\begin{equation}
\begin{split}
  \sum_{m=0}^1 c_me^{i\mu_m(\beta-\tau)}&=\left(\sum_{m=0}^1 c_me^{i\mu_m\tau}\right)^*\\&\approx\sum_{m=0}^1 c_me^{i\mu_m\tau}.  
\end{split}  
\end{equation}
The first equality can be checked using the results of Appendix~\ref{App:c_m_mu_m}, while the approximate equality is appropriate for the slow bath and at high temperatures considered here, when both $\mathcal{C}(t)$ [see Fig.~\ref{Fig:closing}(a)] and $\widetilde{\mathcal{C}}(\tau)$ are almost purely real (the bath is in a good approximation classical).
We note that the exponential decomposition of $\widetilde{\mathcal{C}}(\tau)$ for undamped vibrations~\cite{part1,part2,JChemPhys.159.094113,PhysRevB.105.054311} satisfies the reality and symmetry requirements exactly.

\begin{figure}[htbp!]
    \centering
    \includegraphics[width=\columnwidth]{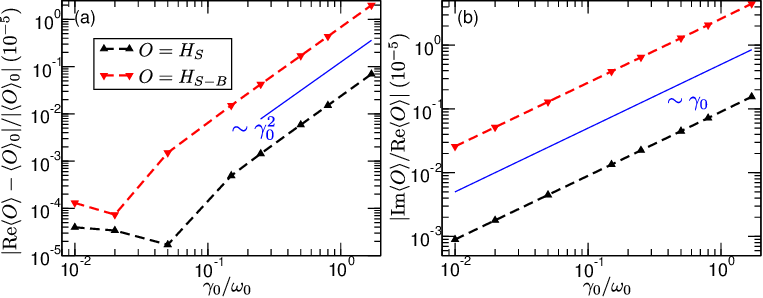}
    \caption{(a) Magnitude of the relative deviation of the real part of the carrier's kinetic energy [Eq.~\eqref{Eq:E_kin}, up-triangles connected by a dashed line] and the carrier--bath interaction energy [Eq.~\eqref{Eq:E_int}, down-triangles connected by a dashed line] from the corresponding values at zero damping (denoted by $\langle O\rangle_0$) as a function of the damping parameter.
    (b) The magnitude of the ratio of the imaginary to the real part of the quantities considered in panel (a) as a function of the damping parameter.
    In both panels, the quantities on vertical axes are given in units of $10^{-5}$, while solid lines show the appropriate power-law scalings.
    The propagation of the i-DEOM [Eq.~\eqref{Eq:im-time-deom}] over the interval $[0,\beta]$ used 1000 imaginary-time steps.}
    \label{Fig:thermodynamics_240625}
\end{figure}

The fact that the symmetries of $\widetilde{\mathcal{C}}(\tau)$ are obeyed only approximately, as well as the existence of various connections between single-carrier levels induced by the hierarchical couplings between DDOs, may suggest that the solution of the i-DEOM in Eq.~\eqref{Eq:im-time-deom} does not provide a faithful representation of the correlated carrier--bath equilibrium.~\cite{JChemPhys.153.154111}
Indeed, while the reduced carrier populations $\langle k|\rho_\mathbf{0}^{(0,\mathrm{eq})}|k\rangle$ are purely real for $\gamma_0=0$,~\cite{part1,part2,JChemPhys.159.094113,PhysRevB.105.054311} we find that the solution of the i-DEOM for $\gamma_0\neq 0$ yields complex populations and complex carrier's kinetic energy
\begin{equation}
\label{Eq:E_kin}
    \langle H_S\rangle=\sum_k\varepsilon_k\left\langle k\left|\rho_\mathbf{0}^{(0,\mathrm{eq})}\right|k\right\rangle.
\end{equation}
In addition, the carrier--bath interaction energy
\begin{equation}
\label{Eq:E_int}
    \langle H_{S-B}\rangle\approx\sum_{kq}M(k,q)\sum_{m=0}^{K-1}\left\langle k\left|\rho_{\mathbf{0}_{qm}^+}^{(1,\mathrm{eq})}\right|k+q\right\rangle
\end{equation}
has a nonzero imaginary part, in contrast to the undamped case.~\cite{part1,part2}
Although Fig.~\ref{Fig:thermodynamics_240625}(a) shows that both $\mathrm{Re}\:\langle H_S\rangle$ and $\mathrm{Re}\:\langle H_{S-B}\rangle$ grow quadratically with $\gamma_0$ (for sufficiently large $\gamma_0$), they are virtually equal to the corresponding values in the undamped case for physically plausible values of $\gamma_0$.
Figure~\ref{Fig:thermodynamics_240625}(b) shows that $\mathrm{Im}\:\langle H_S\rangle$ and $\mathrm{Im}\:\langle H_{S-B}\rangle$ grow linearly with $\gamma_0$ but remain quite small with respect to the corresponding real parts throughout the physically relevant range $\gamma_0/\omega_0\lesssim 2$.
Therefore, the real parts of Eqs.~\eqref{Eq:E_kin} and~\eqref{Eq:E_int} obtained using the i-DEOM remain representative of the kinetic and interaction energies, respectively.

The above-described deficiencies of the DEOM representation of $\rho_\mathrm{tot}^\mathrm{eq}$ following from the i-DEOM can in principle be avoided by propagating the r-DEOM up to very long real times or computing its steady state directly.~\cite{JChemPhys.147.044105}
However, for the parameters considered here, we find that these r-DEOM-based strategies, tried and tested on systems immersed in overdamped baths, perform poorly.

\subsection{Finite-depth and finite-chain effects for weak damping}
\label{SSec:finite_N_D}
\begin{figure*}
    \centering
    \includegraphics[width=\textwidth]{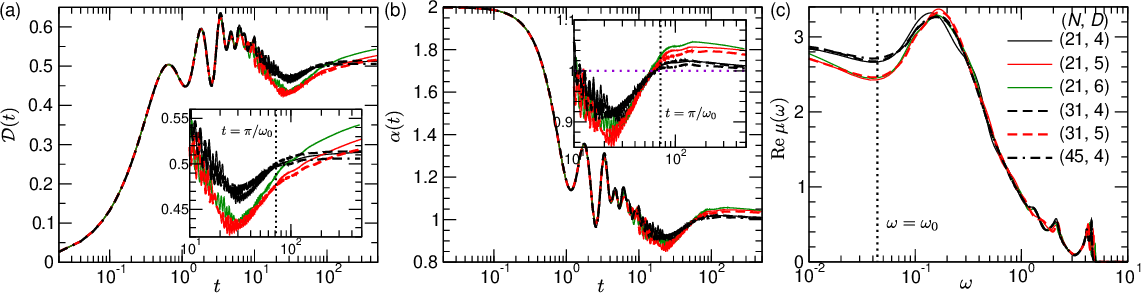}
    \caption{Dynamics of (a) the diffusion constant $\mathcal{D}$ and (b) the diffusion exponent $\alpha$ for different values of the chain length $N$ and maximum hierarchy depth $D$.
    (c) The dynamical-mobility profile $\mathrm{Re}\:\mu(\omega)$ for different values of $N$ and $D$.
    The insets in (a) and (b) zoom in the dynamics of $\mathcal{D}$ and $\alpha$, respectively, on intermediate-to-long timescales.
    The model parameters are $J=1,\omega_0=0.044,\lambda=0.336,T=0.175$, and $\gamma_0/\omega_0=0.05$.
    Note the logarithmic scale on the horizontal axes.}
    \label{Fig:vary_N-D_gamma_0.05_170525}
\end{figure*}

Here, we discuss the maximum DEOM depth $D$ and chain length $N$ needed to obtain transport properties representative of the large-$D$ and large-$N$ limit.

In our recent study,~\cite{part2} which uses $\gamma_0=0$, we have found that $N=31$ is sufficient to keep finite-chain effects under control, while $D=4$ is the largest depth for which the hierarchy closing scheme is effective.
Generally speaking, introducing friction renders finite-size and non-Markovian effects less pronounced, and we expect that the values $N=31$ and $D=4$ remain representative of the large-$N$ and large-$D$ limit also for $\gamma_0\neq 0$.  
However, in Ref.~\onlinecite{part2}, we also concluded that larger values of $D$, required to fully converge $\langle H_S+H_{S-B}\rangle$ with respect to $D$, may be needed to firmly numerically confirm the optical sum rule
\begin{equation}
\label{Eq:OSR}
    \int_0^{+\infty}d\omega\:\mathrm{Re}\:\mu(\omega)=-\frac{\pi}{2}\mathrm{Re}\langle H_S+H_{S-B}\rangle.
\end{equation}
Since a small but nonzero friction generally stabilizes the dynamics, we are now in a position to increase $D$ beyond 4 (at the expense of decreasing $N$), which we do in Figs.~\ref{Fig:vary_N-D_gamma_0.05_170525}(a)--\ref{Fig:vary_N-D_gamma_0.05_170525}(c) for $\gamma_0/\omega_0=0.05$.

Fixing $N=21$ or 31, we observe that transport properties computed for $D=4,5,6$ or $D=4,5$ are qualitatively similar, though there are some quantitative differences.
In particular, the growth of $\mathcal{D}(t)$ beyond $\mathcal{D}(t_\mathrm{min})$ [Fig.~\ref{Fig:vary_N-D_gamma_0.05_170525}(a)], where $t_\mathrm{min}\approx 1/\omega_0$, the overshoot of $\alpha(t)$ above unity on intermediate timescales [Fig.~\ref{Fig:vary_N-D_gamma_0.05_170525}(b)], and the dynamical-mobility enhancement below $\omega_0$ [Fig.~\ref{Fig:vary_N-D_gamma_0.05_170525}(c)] are more pronounced with increasing $D$.
Meanwhile, increasing $D$ does not change either the position or shape of the displaced Drude peak [Fig.~\ref{Fig:vary_N-D_gamma_0.05_170525}(c)], while quantitative changes in transport properties upon increasing $D$ are of the order of $10\%$.~\cite{part2} 
As DEOM computations for $D=5$ (let alone $D=6$) require significantly more computational resources than those employing $D=4$, we opt for $D=4$ as a good compromise between computational effort and accuracy.
The relative accuracy with which the OSR [Eq.~\eqref{Eq:OSR}] is satisfied is $\sim 10^{-4}$ for all depths considered.

Fixing $D=4$ ($D=5$), we find that $\mathcal{D}(t),\alpha(t),$ and $\mathrm{Re}\:\mu(\omega)$ for $N=21$, 31, and 45 ($N=21$ and 31) are qualitatively and to a large extent quantitatively similar.
Finite-chain effects on the dynamics of $\mathcal{D}$ and $\alpha$ come into play for $t\gtrsim\pi/\omega_0$ [see the insets of Figs.~\ref{Fig:vary_N-D_gamma_0.05_170525}(a) and~\ref{Fig:vary_N-D_gamma_0.05_170525}(b)].
We observe that increasing $N$ renders the increase in $\mathcal{D}(t)$ for $t\gtrsim t_\mathrm{min}$, the overshoot of $\alpha(t)$ above unity, and the dynamical-mobility enhancement below $\omega_0$ somewhat less pronounced.
One might thus object to the relevance of these features in the long-chain limit.
To refute such an objection, we note that the very same features have been observed in fully quantum~\cite{mitric2024dynamicalquantumtypicalitysimple} or quantum--classical~\cite{mitric2024precursorsandersonlocalizationholstein,runeson2024,PhysRevX.10.021062,PhysRevB.107.064304} approaches to carrier transport in models with local~\cite{mitric2024dynamicalquantumtypicalitysimple,mitric2024precursorsandersonlocalizationholstein} or nonlocal~\cite{runeson2024,PhysRevX.10.021062,PhysRevB.107.064304} interaction of a carrier with undamped~\cite{mitric2024dynamicalquantumtypicalitysimple,mitric2024precursorsandersonlocalizationholstein,runeson2024,PhysRevX.10.021062,PhysRevB.107.064304} or moderately damped~\cite{runeson2024} vibrations.
In particular, the quantum--classical study in Ref.~\onlinecite{runeson2024} considered the same adiabaticity $\omega_0/J$ ratio as here and observed the same long-time (low-frequency) features in $\mathcal{D}(t)$ [$\mathrm{Re}\:\mu(\omega)$] on chains with hundreds of sites.
By assuming classical vibrational dynamics, which is appropriate for small $\omega_0/J$ and at the high temperature considered here, quantum--classical methods manage to reach what would be the $D\to\infty$ limit in the DEOM language.
Keeping in mind that increasing $D$ beyond 4 renders the long-time (low-frequency) features more pronounced, their existence for small $\gamma_0$ can be considered as firmly established.

The above observations strengthen the conclusions we reached in the dissipationless case,~\cite{part2} firmly establishing the transient nature of the super-to-subdiffusive crossover and the low-frequency enhancement in the carrier's optical response in the case of weakly damped vibrations moderately coupled to the carrier. 

\section{Results}
\label{Sec:results}
\subsection{Influence of damping on transport dynamics}
\label{SSec:TLS_vs_DEOM}
\begin{figure*}
    \centering
    \includegraphics[width=\textwidth]{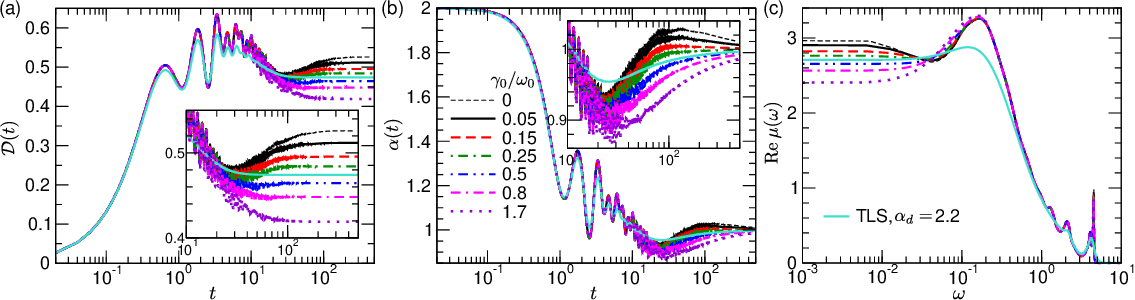}
    \caption{Dynamics of (a) the diffusion constant $\mathcal{D}$ and (b) the diffusion exponent $\alpha$ for different values of the friction coefficient $\gamma_0$.
    (c) The dynamical-mobility profile $\mathrm{Re}\:\mu(\omega)$ for different values of $\gamma_0$.
    The TLS predictions employ $\alpha_d=2.2$.
    The insets in (a) and (b) display the dynamics of $\mathcal{D}$ and $\alpha$, respectively, on intermediate-to-long timescales.
    The model parameters are $J=1,\omega_0=0.044,\lambda=0.336,$ and $T=0.175$.
    DEOM computations are performed for $N=31,D=4$.
    Note the logarithmic scale on the horizontal axes.
    }
    \label{Fig:vary_gamma0_N_31_D_4_170525}
\end{figure*}

In Figs.~\ref{Fig:vary_gamma0_N_31_D_4_170525}(a)--\ref{Fig:vary_gamma0_N_31_D_4_170525}(c), we monitor the changes to transport quantities brought about by the nonzero width of the spectrum of relevant phonons, which is embodied in the damping parameter $\gamma_0$ of the present model [Eq.~\eqref{Eq:J_BO_w}].

Figures~\ref{Fig:vary_gamma0_N_31_D_4_170525}(a) and~\ref{Fig:vary_gamma0_N_31_D_4_170525}(b) reveal that a realistically strong damping of vibrational motions starts to influence carrier dynamics on intermediate timescales (for $t\gtrsim t_\mathrm{min}\approx 1/\omega_0$).
Therefore, realistically strong damping (a realistically wide phonon spectrum) does not change the position and weakly affects the shape of the displaced Drude peak, which reflects short-time carrier dynamics in the field of effectively frozen vibrations~\cite{mitric2024precursorsandersonlocalizationholstein,AdvFunctMater.26.2292,PhysRevB.83.081202,PhysRevLett.132.186303} [see Fig.~\ref{Fig:vary_gamma0_N_31_D_4_170525}(c)].
The timescale $\tau_\mathrm{diff}$ on which the diffusive carrier dynamics is established can be estimated as the characteristic timescale for the decay of the current autocorrelation function, which can be defined as [see also Eq.~\eqref{Eq:def_tau_c}]
\begin{equation}
\label{Eq:def_tau_diff}
    \tau_\mathrm{diff}=\int_0^{+\infty}dt\:\frac{\mathrm{Re}\:C_{jj}(t)}{\mathrm{Re}\:C_{jj}(0)}=\frac{\mathcal{D}_{\infty}}{\langle j^2\rangle}.
\end{equation}
Equation~\eqref{Eq:def_tau_diff} involves the long-time diffusion constant, which somewhat decreases with damping~\cite{JChemPhys.160.111102} [see the inset of Fig.~\ref{Fig:vary_gamma0_N_31_D_4_170525}(a)], and the thermodynamic quantity $\langle j^2\rangle$, which is expected to be virtually independent of the damping (see Sec.~\ref{SSec:thermodynamics_OSR}).
While the approach toward diffusive transport becomes somewhat more rapid with increasing damping, the order of magnitude of $\tau_\mathrm{diff}$ remains constant for realistic values of $\gamma_0$.

Increasing $\gamma_0$ renders the intermediate-time growth of $\mathcal{D}(t)$ less pronounced until it is fully suppressed for $\gamma_0/\omega_0\gtrsim 0.5$ [see the inset of Fig.~\ref{Fig:vary_gamma0_N_31_D_4_170525}(a)].
Interestingly, the magnitude of the growth, quantified by $\mathcal{D}_\infty-\mathcal{D}(t_\mathrm{min})$, determines whether the super-to-subdiffusive crossover is transient or permanent [see Fig.~\ref{Fig:vary_gamma0_N_31_D_4_170525}(b)] as well as the low-frequency features of the dynamical mobility [see Fig.~\ref{Fig:vary_gamma0_N_31_D_4_170525}(c)].
For $\gamma_0/\omega_0=0.05$ and 0.15, $\mathcal{D}_\infty-\mathcal{D}(t_\mathrm{min})$ is sufficiently large so that the diffusive transport sets in from the superdiffusive side, giving rise to the enhancement in the dynamical mobility below the vibrational frequency.
Although $\mathcal{D}_\infty-\mathcal{D}(t_\mathrm{min})$ remains positive for $\gamma_0/\omega_0=0.25$, we find that its magnitude is such that the diffusive transport is approached from the subdiffusive side, while there is almost no low-frequency enhancement in the dynamical mobility [see also Fig.~\ref{Fig:enaqt}(b)].

\subsection{Success of the transient localization phenomenology}
Notably, the approach to diffusive dynamics from the subdiffusive side and the absence of dynamical-mobility enhancements below phonon frequency, which persist for $\gamma_0/\omega_0\gtrsim 0.25$, coincide with the predictions of the TLS.~\cite{PhysRevB.83.081202,AdvFunctMater.26.2292}
The TLS starts from the frozen-phonon limit, as appropriate on short timescales $t\ll\omega_0^{-1}$, and assumes that the long-time carrier diffusion, absent in the frozen-phonon limit,~\cite{PhysRev.109.1492} is reached through effective carrier--phonon dynamics.
In the relaxation-time approximation,~\cite{PhysRevB.83.081202,AdvFunctMater.26.2292} this amounts to
\begin{equation}
\label{Eq:TLS-RTA}
    C_{jj}^\mathrm{TLS}(t)=C_{jj}^\mathrm{dis}(t)\:e^{-|t|/\tau_d},
\end{equation}
where $C_{jj}^\mathrm{dis}(t)$ is the current autocorrelation function of the corresponding model with static disorder in nearest-neighbor hopping amplitudes.
Rewriting $H_{S-B}$ in Eq.~\eqref{Eq:def_H_tot_momentum} in real space,
\begin{equation}
\label{Eq:def_H_S_B_real_space}
\begin{split}
    H_{S-B}=&\sum_n(|n\rangle\langle n+1|+|n+1\rangle\langle n|)\times\\&\sum_\xi g_\xi\left(b_{n\xi}^\dagger+b_{n\xi}-b_{n+1,\xi}^\dagger-b_{n+1,\xi}\right),
\end{split}
\end{equation}
the statically disordered model is constructed by formally replacing the phonon operator
$$\sum_\xi g_\xi\left(b_{n\xi}^\dagger+b_{n\xi}-b_{n+1,\xi}^\dagger-b_{n+1,\xi}\right)$$
by the Gaussian random variable $X_{n,n+1}$ of zero mean and variance
\begin{equation}
\label{Eq:def_sigma_squared}
    \sigma^2=\sum_\xi 2\frac{2g_\xi^2}{\beta\omega_\xi}=2\lambda JT.
\end{equation}
We have used that the expectation value of the operator $g_\xi(b_{n\xi}^\dagger+b_{n\xi})$ is distributed according to the Gaussian distribution of zero mean and variance $g_\xi^2\coth(\beta\omega_\xi/2)\approx 2g_\xi^2/(\beta\omega_\xi)$.~\cite{mitric2024precursorsandersonlocalizationholstein}
The approximate equality holds for $\beta\omega_\xi\ll 1$, as appropriate in the high-temperature slow-phonon regime (see Sec.~\ref{SSec:model_parameters}) and on short timescales we consider.
As the operators $g_\xi(b_{n\xi}^\dagger+b_{n\xi})$ commute for different values of $n$ or $\xi$, the expectation value of their sum over $\xi$ is distributed in the same way as the sum of mutually independent Gaussian variables, each of which has zero mean and variance $2g_\xi^2/(\beta\omega_\xi)$, thus justifying Eq.~\eqref{Eq:def_sigma_squared}.
Equation~\eqref{Eq:def_sigma_squared} additionally reveals that the reference statically disordered model does not depend on the details of the carrier--phonon SD or phonon dynamics [Eq.~\eqref{Eq:def_mathcalC_t}] but only on the overall interaction strength quantified by the dimensionless parameter $\lambda$ [Eq.~\eqref{Eq:def_lambda}].
This once again reflects the fact that the statically disordered model is appropriate only on short timescales. 

In Eq.~\eqref{Eq:TLS-RTA}, the timescale $\tau_d$, after which phonon dynamics cannot be neglected, is assumed as $\tau_d^{-1}=\alpha_d\omega_0$,~\cite{PhysRevB.83.081202,AdvFunctMater.26.2292} where the free parameter $\alpha_d\sim 1$ can be finetuned to best reproduce some reference (experimental or numerical) result for $\mu_\mathrm{dc}$ without qualitatively affecting TLS predictions for transport dynamics.
In Figs.~\ref{Fig:vary_gamma0_N_31_D_4_170525}(a)--\ref{Fig:vary_gamma0_N_31_D_4_170525}(c), we show the TLS result using $\alpha_d=2.2$.
This value was suggested by comparing TLS and reference numerically exact results in Refs.~\onlinecite{PhysRevResearch.2.013001,PhysRevLett.132.266502}, which considered undamped vibrations.
Here, this standard choice for $\alpha_d$ gives $\mu_\mathrm{dc}^\mathrm{TLS}$ that differs by at most $20\%$ from the corresponding DEOM results for physically reasonable damping strengths [see the inset of Fig.~\ref{Fig:vary_gamma0_N_31_D_4_170525}(a)].
Importantly, both the TLS with $\alpha_d=2.2$ and DEOM with physically relevant values of $\gamma_0$ yield $\mu_\mathrm{dc}$ that falls into the experimental range $\mu_\mathrm{dc}^\mathrm{exp}\sim (10-20)\:\mathrm{cm}^2/(\mathrm{Vs})$ for the room-temperature dc mobility in rubrene~\cite{PhysRevLett.93.086602,PhysRevLett.95.226601,PhysRevLett.98.196804} [our DEOM mobility is measured in units of $e_0a_l^2/\hbar$, corresponding to $7.9\:\mathrm{cm}^2/(\mathrm{Vs})$ for $a_l=7.2\mathrm{\AA}$, see, e.g., Ref.~\onlinecite{AdvMater.19.2000}].

Therefore, Figs.~\ref{Fig:vary_gamma0_N_31_D_4_170525}(a)--\ref{Fig:vary_gamma0_N_31_D_4_170525}(c) strongly suggest that a nonzero width of the spectrum of relevant phonons,~\cite{AdvMater.19.2000,JPhysChemA.110.4065,JChemPhys.129.094506} which in the present model amounts to a nonzero damping in intermolecular motions,~\cite{PhysRevLett.102.116602,JChemPhys.131.014703,ChemPhysChem.11.2067} is behind the success of the transient-localization phenomenology (and the related Drude--Anderson framework) in explaining the experimental carriers' optical response.
Notably, the damping does not have to be excessively strong (the relevant-phonon spectrum does not have to be excessively wide) in order for DEOM results to comply with the transient-localization phenomenology.
Figures~\ref{Fig:vary_gamma0_N_31_D_4_170525}(a)--\ref{Fig:vary_gamma0_N_31_D_4_170525}(c) reveal that this happens already for $\gamma_0/\omega_0\gtrsim 0.25$, which belongs to the underdamped regime characterized by a pronounced peak in the SD (see Fig.~\ref{Fig:fig_sd_200525}).
We thus propose that the optical-absorption enhancements below the vibrational frequency reported in model-Hamiltonian studies~\cite{part2,mitric2024dynamicalquantumtypicalitysimple,mitric2024precursorsandersonlocalizationholstein,runeson2024,PhysRevX.10.021062,PhysRevB.107.064304} are to be regarded as artifacts of the delta-like phonon spectrum (undamped vibrations).

\subsection{A closer look into the TLS phenomenological timescale}

\begin{figure}[htbp!]
    \centering
    \includegraphics[width=0.9\columnwidth]{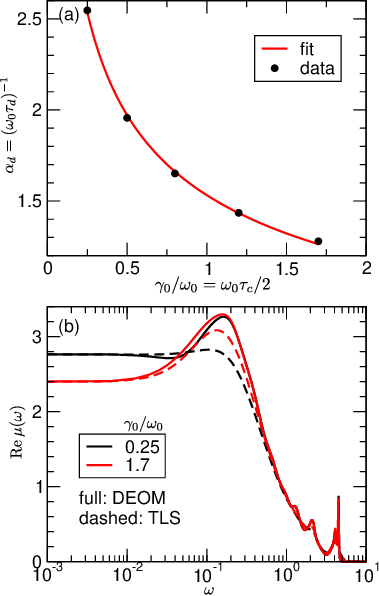}
    \caption{(a) TLS parameter $\alpha_d=(\omega_0\tau_d)^{-1}$ [Eq.~\eqref{Eq:TLS-RTA}], chosen so that $\mu_\mathrm{dc}^\mathrm{TLS}=\mu_\mathrm{dc}^\mathrm{DEOM}$, as a function of $\gamma_0/\omega_0=\omega_0\tau_c/2$ [Eq.~\eqref{Eq:def_tau_c}].
    The best power-law fit $\alpha_d=a_0(\gamma_0/\omega_0)^{-a_1}$, with $a_0=1.53$ and $a_1=0.364$, of the data (dots) is shown as a solid line.
    (b) The comparison of DEOM (solid lines) and TLS [with $\alpha_d$ chosen as in panel (a), dashed lines] dynamical-mobility profiles for $\gamma_0/\omega_0=0.25$ (black) and 1.7 (red).
    }
    \label{Fig:enaqt}
\end{figure}

The parameter $\tau_d$ in Eq.~\eqref{Eq:TLS-RTA} is associated with the characteristic timescale $\tau_c$ of phonon-induced fluctuations of the transfer integral.
In models with undamped phonons and in the limit of zero carrier density, it exhibits undamped oscillations with phonon frequency $\omega_0$~\cite{JPhysCondensMatter.18.7299,PhysRevB.105.054311} so that $\tau_c\propto\omega_0^{-1}$.
However, the fact that Eq.~\eqref{Eq:TLS-RTA} features an exponentially decaying (and not an oscillatory) factor suggests that the TLS is even more appropriate for dissipative models, in which the lifetime of transfer-integral fluctuations is finite.
In our model, $\tau_c$ can be estimated as~\cite{JPhysSocJpn.89.015001}
\begin{equation}
\label{Eq:def_tau_c}
    \tau_c=\int_0^{+\infty}dt\:\frac{\Psi(t)}{\Psi(0)}=2\frac{\gamma_0}{\omega_0^2},
\end{equation}
where $\Psi(t)=-2\int_t^{+\infty}ds\:\mathrm{Im}\:\mathcal{C}(s)$ is the so-called relaxation function.~\cite{Kubo-noneq-stat-mech-book,JPhysSocJpn.89.015001}

For $\gamma_0/\omega_0\gtrsim 0.25$, for which $\alpha(t)$ and $\mathrm{Re}\:\mu(\omega)$ display TL phenomenology, we extract $\alpha_d=(\omega_0\tau_d)^{-1}$ by requiring that the dc mobility within the TLS is identical to the corresponding DEOM result.
Figure~\ref{Fig:enaqt}(a) shows that the extracted $\alpha_d$ is correlated with $\tau_c$, see the powerlaw fit (solid line) whose parameters are summarized in the caption.
Interestingly, $\tau_d=(\alpha_d\omega_0)^{-1}$ somewhat increases with the damping, without changing its order of magnitude, contrary to the timescale $\tau_\mathrm{diff}$ on which carrier dynamics become diffusive (see Sec.~\ref{SSec:TLS_vs_DEOM}).  
[Since we choose $\tau_d$ so that $\mu_\mathrm{dc}^\mathrm{TLS}=\mu_\mathrm{dc}^\mathrm{DEOM}$, and since $C_{jj}(0)$ is identical within TLS and DEOM, Eq.~\eqref{Eq:def_tau_diff} shows that $\tau_\mathrm{diff}$ is also identical within TLS and DEOM.]
Such a trend in $\tau_d$ with increasing $\gamma_0$ is compatible with the dc mobility decreasing with $\gamma_0$, see Figs.~\ref{Fig:vary_gamma0_N_31_D_4_170525}(a) and~\ref{Fig:vary_gamma0_N_31_D_4_170525}(c).
In particular, as vibrations become more damped, the model becomes increasingly similar to that with only static disorder,~\cite{JChemPhys.131.014703} and this reference point of the TLS is gradually approached as $\tau_d$ is increased [see Eq.~\eqref{Eq:TLS-RTA}].
This is discussed in more detail in Appendix~\ref{App:strong_damping}.

If we choose $\alpha_d$ so that the TLS perfectly reproduces DEOM mobility, the overall agreement between TLS and DEOM dynamical-mobility profiles improves with increasing $\gamma_0$ [see Fig.~\ref{Fig:enaqt}(b)].
At smaller damping strengths, the requirement $\mu_\mathrm{dc}^\mathrm{TLS}=\mu_\mathrm{dc}^\mathrm{DEOM}$ makes the TLS almost completely miss the DDP [see the curves for $\gamma_0/\omega_0=0.25$ in Fig.~\ref{Fig:enaqt}(b)].
Although a more accurate TLS approximation to the numerically exact DDP would underestimate the dc mobility, $\mu_\mathrm{dc}^\mathrm{TLS}$ would still be of the same order of magnitude as $\mu_\mathrm{dc}^\mathrm{DEOM}$.
Meanwhile, at larger damping strengths, the TLS can reproduce reasonably well both $\mu_\mathrm{dc}$ and the position, intensity, and shape of the DDP [see the curves for $\gamma_0/\omega_0=1.7$ in Fig.~\ref{Fig:enaqt}(b)].
For even stronger dampings, we expect that the dependence of $\tau_d$ on $\gamma_0$ is different from that in Fig.~\ref{Fig:enaqt}(a).
In Appendix~\ref{App:strong_damping}, we argue that $\tau_d$ should depend linearly on $\gamma_0$ in the strong-damping limit.

\subsection{Transport dynamics for the Drude--Lorentz spectral density}
\label{SSec:transport-DL}

\begin{figure}
    \centering
    \includegraphics[width=\columnwidth]{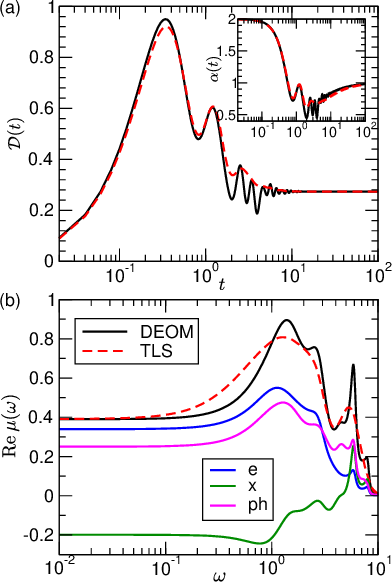}
    \caption{Time-dependent diffusion constant $\mathcal{D}(t)$ (a) and the dynamical-mobility profile $\mathrm{Re}\:\mu(\omega)$ (b) computed assuming the DL SD and using the DEOM (solid line) and TLS (dashed line).
    The TLS parameter $\alpha_d=(\gamma_\mathrm{DL}\tau_d)^{-1}=1.66$ is chosen so that $\mu_\mathrm{dc}^\mathrm{DEOM}=\mu_\mathrm{dc}^\mathrm{TLS}$.
    The inset of panel (a) displays the dynamics of the diffusion exponent $\alpha$.
    Panel (b) also shows the purely electronic (label "e"), phonon-assisted (label "ph"), and cross (label "x") contributions to $\mathrm{Re}\:\mu(\omega)$.
    The model parameters are $J=1,\gamma_\mathrm{DL}=2/15,\lambda=0.5,T=0.7$.}
    \label{Fig:drude_lorentz}
\end{figure}

For the DL SD [Eq.~\eqref{Eq:J_DL_omega}], the short-time behavior of the free-bath correlation function [Eq.~\eqref{Eq:def_mathcalC_t}] is unphysical because its imaginary part~\cite{Mukamel-book,Valkunas-Abramavicius-Mancal-book}
\begin{equation}
\begin{split}
    \mathrm{Im}\:\mathcal{C}_\mathrm{DL}(t)=-E_0\gamma_\mathrm{DL}e^{-\gamma_\mathrm{DL}t}
\end{split}
\end{equation}
does not vanish at $t=0$ as it should.~\cite{JPhysSocJpn.89.015001,JChemPhys.130.234111}
The fact that $\mathrm{Im}\:\mathcal{C}_\mathrm{DL}(0)\neq 0$ is connected to the long tail of $\mathcal{J}_\mathrm{DL}(\omega)$ in the high-frequency region, and underscores the coarse-grained nature of the short-time dynamics.~\cite{JPhysSocJpn.89.015001,JChemPhys.130.234111}
This drawback of the DL SD is not expected to appreciably affect carrier transport dynamics inferred from $C_{jj}(t)$ in Holstein-like models,~\cite{JChemPhys.142.174103,JChemPhys.150.234101,JChemTheoryComput.20.7052,ProcNatlAcadSci.122.e2424582122,ChemSci.15.16715} in which there is no phonon-assisted current.
Meanwhile, in models with nonlocal carrier--phonon interaction, the coarse-grained nature is reflected in the unphysical initial condition $\langle j_\mathrm{e-ph}^2\rangle$ for the phonon-assisted contribution $\langle j_\mathrm{e-ph}(t)j_\mathrm{e-ph}(0)\rangle$ to $C_{jj}(t)$, which we discuss in more detail in Appendix~\ref{App:coarse-grained-DL}.
This issue is avoided through the procedure by Ishizaki,~\cite{JPhysSocJpn.89.015001} which also helps us keep the number of exponentially decaying terms in Eq.~\eqref{Eq:def_mathcalC_t} low.

Considering the DL SD, the model parameters are usually chosen as in Ref.~\onlinecite{JChemPhys.132.081101}, so that $J=300\:\mathrm{cm}^{-1}$, $2E_0=323\:\mathrm{cm}^{-1}$, and $\gamma_\mathrm{DL}=41\:\mathrm{cm}^{-1}$.
However, the models considered in Refs.~\onlinecite{JChemPhys.132.081101,JChemPhys.161.084118} somewhat differ from the model we consider.
In particular, the nearest-neighbor hopping amplitude in Ref.~\onlinecite{JChemPhys.132.081101} (Ref.~\onlinecite{JChemPhys.161.084118}) is modulated by the displacement of the oscillator on the left end of a bond (the center of a bond) so that the dimensionless interaction constant is $\lambda=\frac{E_0}{J}$.
This is in contrast to our definition $\lambda=\frac{2E_0}{J}$, which reflects the fact that the hopping amplitude is modulated by the difference of the displacements on two ends of a bond.
Importantly, $\lambda\approx 0.5$ in Refs.~\onlinecite{JChemPhys.132.081101,JChemPhys.161.084118}, which belongs to the intermediate-interaction regime relevant for organic molecular crystals.~\cite{JChemPhys.152.190902}

Figures~\ref{Fig:drude_lorentz}(a) and~\ref{Fig:drude_lorentz}(b) present DEOM results for Ishizaki's BO approximation to the DL SD~\cite{JPhysSocJpn.89.015001} assuming that $J=300\:\mathrm{cm}^{-1}$, $E_0=75\:\mathrm{cm}^{-1}$ ($\lambda=0.5$), $\gamma_0=2\gamma_\mathrm{DL}=80\:\mathrm{cm}^{-1}$ ($\gamma_0/J=4/15$), and $\epsilon=\gamma_0/4=20\:\mathrm{cm}^{-1}$ ($\gamma_0/\omega_0=4/\sqrt{17}$), at room temperature $T=210\:\mathrm{cm}^{-1}$ ($T/J=0.7$).
DEOM computations are performed using $N=11$, $D=8$, and $K=2$.
We have checked that increasing $D$ from 8 to 9 or retaining more than two terms in the exponential decomposition of Eq.~\eqref{Eq:def_mathcalC_t} ($K=3$ or 4 with $N=7$) introduces only minor quantitative changes to the DEOM result for $N=11,D=8,K=2$.
We have also checked that employing a smaller value of $\epsilon=\gamma_0/20$ does not change (either qualitatively or quantitatively) the results using $\epsilon=\gamma_0/4$.

The DEOM results in Figs.~\ref{Fig:drude_lorentz}(a) and~\ref{Fig:drude_lorentz}(b) can be well reproduced by the TL phenomenology using the phenomenological timescale $\tau_d^{-1}=\alpha_d\gamma_\mathrm{DL}$ with $\alpha_d\sim 1$.
Requiring that $\mu_\mathrm{dc}^\mathrm{DEOM}=\mu_\mathrm{dc}^\mathrm{TLS}$, we obtain $\alpha_d=1.66$.
The good performance of the exponential ansatz in Eq.~\eqref{Eq:TLS-RTA} for the DL SD is connected to the corresponding bath correlation function containing only exponentially decaying terms, the most dominant one being proportional to $e^{-\gamma_\mathrm{DL}t}$.

Finally, for the usual choice of model parameters of the DL SD at room temperature,~\cite{JChemPhys.132.081101,JChemPhys.161.084118} the phonon-assisted contribution to $\mu_\mathrm{dc}$ [obtained by considering only $\langle j_\mathrm{e-ph}(t)j_\mathrm{e-ph}(0)\rangle$ in Eq.~\eqref{Eq:def_re_mu_w}] is comparable to the purely electronic contribution to $\mu_\mathrm{dc}$ [obtained by considering only $\langle j_\mathrm{e}(t)j_\mathrm{e}(0)\rangle$ in Eq.~\eqref{Eq:def_re_mu_w}] [see Fig.~\ref{Fig:drude_lorentz}(b)].
This is different from the situation for model parameters we use with BO SD at room temperature, for which the phonon-assisted contribution is much smaller than the purely electronic contribution to $\mu_\mathrm{dc}$,~\cite{part2} irrespective of the damping analyzed.
The reason behind this difference lies in the different values of the transfer integral considered in the two models.
In particular, the authors of Ref.~\onlinecite{JChemPhys.132.081101} suggest that the Holstein-type interaction with high-frequency intramolecular vibrations leads to an effective reduction of the hopping amplitude from its bare value $J\approx 1150\:\mathrm{cm}^{-1}$ (see, e.g., Ref.~\onlinecite{AdvMater.19.2000}) to the dressed value $J\approx 300\:\mathrm{cm}^{-1}$.
Keeping this in mind and remembering that the value of $\lambda$ used with the DL SD (0.5) is somewhat larger than that used with the BO SD (0.336), it is not surprising that the dc mobility in the parameter regime discussed in Refs.~\onlinecite{JChemPhys.132.081101,JChemPhys.161.084118} is almost an order of magnitude smaller than in the regime analyzed in Fig.~\ref{Fig:vary_gamma0_N_31_D_4_170525} [compare Fig.~\ref{Fig:drude_lorentz}(b) to Fig.~\ref{Fig:vary_gamma0_N_31_D_4_170525}(c)].

\section{Summary and outlook}
\label{Sec:summary-outlook}
In this study, we have considered an underexplored aspect of carrier transport in molecular semiconductors, namely, how a nonzero width of the spectrum of the vibrations modulating carrier's hopping amplitude influences transport dynamics.
Previous studies considered this aspect using the Drude--Lorentz spectral density and inferred transport properties from carrier dynamics starting from the factorized initial condition.
Our considerations are instead based on the numerically ``exact'' quantum dynamics of the current autocorrelation function of a one-dimensional model with the physically more plausible (and formally more appropriate) Brownian-oscillator spectral density of nonlocal carrier--vibrational interaction.
We use the DEOM representation of quantum dynamics, whose dissipaton algebra is essential to correctly treat the phonon-assisted current.

Whenever weakly damped (or undamped) vibrational dynamics are taken seriously, either numerically ``exactly'' or approximately, the diffusive carrier dynamics in the physically relevant high-temperature intermediate-interaction regime set in from the superdiffusive side, giving rise to enhancements in the dynamical mobility below the vibrational frequency.
Considering damping strengths (widths of the relevant phonon spectrum) characteristic of realistic materials, we find that the diffusive transport is approached from the subdiffusive side and that the dynamical mobility exhibits no low-frequency upturns.
Notably, this is in qualitative agreement with the majority of experimental results and the widely used transient localization scenario.
Our findings explain the success of the exponentially decaying ansatz underlying the TLS in rationalizing experimental data.
We suggest that low-frequency enhancements in the dynamical-mobility profiles in minimal models reflect the widely used assumption of a delta-like phonon spectrum (undamped vibrations).

On the theoretical side, our study represents one of the pioneering applications of the DEOM formalism to computations of finite-temperature real-time correlation functions in dissipative electron--phonon models with more than a few electronic levels and nonlocal electron--phonon interaction.
We have also discussed the pitfalls of the widely used Drude--Lorentz bath model, which become apparent when considering mixed system--bath quantities, and how to avoid them. 
Thanks to the physically justified high-temperature assumption, our momentum-space formulation, and an appropriate DEOM closing with respect to the maximum depth $D$, we have succeeded in obtaining physically relevant results without resorting to hierarchy representations in nonstandard basis sets or using advanced DEOM propagation techniques.
These could be exploited in further investigations of more realistic and challenging models, such as the multi-mode (instead of the single-mode) Brownian-oscillator model for the nonlocal interaction of the carrier with low-frequency intermolecular phonons or the explicit inclusion of the local interaction with high-frequency intramolecular vibrations.

\acknowledgments
This research was supported by the Science Fund of the Republic of Serbia, Grant No. 5468, Polaron Mobility in Model Systems and Real Materials--PolMoReMa.
The author also acknowledges funding provided by the Institute of Physics Belgrade through a grant from the Ministry of Science, Technological Development, and Innovation of the Republic of Serbia.
Numerical computations were performed on the PARADOX-IV supercomputing facility at the Scientific Computing Laboratory, National Center of Excellence for the Study of Complex Systems, Institute of Physics Belgrade.
The author thanks Nenad Vukmirovi\'c for the careful reading of the manuscript.

\section*{Data availability}
The data that support the findings of this study are available from the corresponding author upon reasonable request.

\appendix
\section{Coefficients $c_m$ and $\mu_m$ in Eq.~\eqref{Eq:def_mathcalC_t} for the Brownian-oscillator spectral density}
\label{App:c_m_mu_m}
In Eq.~\eqref{Eq:def_mathcalC_t}, the terms with $m=0$ and 1 stem from the poles of $\mathcal{J}(\omega)$ and read~\cite{JPhysChemLett.10.2665,JChemPhys.160.111102}
\begin{equation}
\label{Eq:A1}
\begin{split}
    \mu_m=\gamma_0+(-1)^m i\zeta,
\end{split}
\end{equation}
\begin{equation}
\label{Eq:A2}
    c_m=(-1)^{m+1}\frac{E_0\omega_0^2}{2\zeta}\left[\coth\left(\frac{i\beta\mu_m}{2}\right)-1\right],
\end{equation}
where
\begin{equation}
    \zeta=\begin{cases}
        &\sqrt{\omega_0^2-\gamma_0^2},\:\gamma_0<\omega_0\\
        &i\sqrt{\gamma_0^2-\omega_0^2},\:\gamma_0>\omega_0.
    \end{cases}
\end{equation}

The terms of Eq.~\eqref{Eq:def_mathcalC_t} with $m\geq 2$ are connected to the decomposition
\begin{equation*}
    \frac{1}{1-e^{-z}}=\frac{1}{2}+\frac{1}{z}+\sum_{m=1}^{+\infty}\rho_m\left(\frac{1}{z-i\xi_m}+\frac{1}{z+i\xi_m}\right)
\end{equation*}
of the Bose--Einstein function in terms of its simple poles ($0,\pm i\xi_m$) and residues ($1,\rho_m$).
Therefore, for $m\geq 2$,~\cite{JPhysChemLett.10.2665,JChemPhys.160.111102}
\begin{equation}
    \mu_m=\xi_{m-1}T,
\end{equation}
\begin{equation}
    \frac{c_m}{E_0\omega_0}=\frac{-8\xi_{m-1}\rho_{m-1}(\beta\omega_0)(\beta\gamma_0)}{[(\beta\omega_0)^2+\xi_{m-1}^2]^2-4\xi_{m-1}^2(\beta\gamma_0)^2}.
\end{equation}
In the Matsubara decomposition, $\xi_m=2\pi m,\rho_m=1$.
In actual computations, we follow Ref.~\onlinecite{JChemPhys.141.094101} and obtain $\xi_m$ and $\rho_m$ using the $(N_\mathrm{BE}-1)/N_\mathrm{BE}$ Pad\'e approximant to the Bose--Einstein function.~\cite{JChemPhys.133.101106,JChemPhys.134.244106}

\section{Strong-damping limit}
\label{App:strong_damping}
Here, we discuss the bath correlation function $\mathcal{C}(t)$ in the strong-damping limit $\gamma_0\gg\omega_0$. 
Equation~\eqref{Eq:A1} determining the rates of the exponentials in Eq.~\eqref{Eq:def_mathcalC_t} then becomes
\begin{equation}
\label{Eq:BB1}
    \mu_0\approx\frac{\omega_0^2}{2\gamma_0},\:\mu_1\approx 2\gamma_0,
\end{equation}
where we keep only the leading terms.
The term $e^{-\mu_0t}$ tends to unity with increasing $\gamma_0$, which is a formal manifestation of the system gradually becoming similar to the statically disordered one, \emph{vide infra}.
Meanwhile, the term $e^{-\mu_1t}$ becomes increasingly similar to a (one-sided) $\delta$ function, and the same holds for Matsubara terms at sufficiently high temperatures.
In other words, in the strong-damping limit and at high temperatures,
\begin{equation}
\label{Eq:BB2}
    \mathcal{C}(t)\approx c_0+2\Delta\delta(t),
\end{equation}
where $\Delta$ is given by Eq.~\eqref{Eq:def_delta} with $K=0$.
Using Eq.~\eqref{Eq:A2} for $m=0$ and Eqs.~\eqref{Eq:BB1} and~\eqref{Eq:def_delta} and keeping only the leading-order terms, we recast Eq.~\eqref{Eq:BB2} as
\begin{equation}
\label{Eq:BB3}
    \mathcal{C}(t)\approx 2E_0T-\frac{4E_0T}{\gamma_0}\delta(t).
\end{equation}

By representing Gaussian static disorder in terms of appropriate bosonic operators, Ref.~\onlinecite{JChemPhys.155.134102} formally establishes that quantum dynamics of a carrier in the bath defined by Eq.~\eqref{Eq:BB3} can be computed by averaging the dynamics modulated by the white noise in transfer integrals [the second term in Eq.~\eqref{Eq:BB3}] over different realizations of Gaussian static disorder in transfer integrals [the first term in Eq.~\eqref{Eq:BB3}].
$\mathcal{C}(t)$ in Eq.~\eqref{Eq:def_mathcalC_t} and its approximate version Eq.~\eqref{Eq:BB3} is the single-site bath correlation function, while nearest-neighbor transfer integrals are modulated by the difference between displacements of the corresponding local oscillators [Eq.~\eqref{Eq:def_H_S_B_real_space}].
Therefore, the disorder strength is two times greater than in Eq.~\eqref{Eq:BB3}, i.e., $\sigma^2=2\times 2E_0T=2\lambda JT$ [Eq.~\eqref{Eq:def_lambda}], which coincides with Eq.~\eqref{Eq:def_sigma_squared}.
The white-noise term ensures that the long-time diffusion constant remains nonzero for large but finite $\gamma_0$, which is nowadays known as the environment-assisted (or noise-assisted) quantum transport.~\cite{NewJPhys.11.033003,JChemPhys.131.105106,NewJPhys.15.085010}
The TLS avoids computing (and then averaging) noise-modulated dynamics in many static-disorder realizations by resorting to Eq.~\eqref{Eq:TLS-RTA}, which approximates noise effects through the exponential correction to the dynamics in the static landscape.

The approximation embodied in Eq.~\eqref{Eq:TLS-RTA} provides the exact solution to white noise-modulated transport dynamics in the ordered Holstein model if we set $\tau_d=\gamma_0/(2E_0T)$ and $C_{jj}^\mathrm{dis}(t)=2J^2$.~\cite{PhysRevLett.39.1424,NewJPhys.15.085010}
Although the exact solution to white noise-modulated transport dynamics in the ordered Peierls model is more complicated,~\cite{PhysRevLett.39.1424} exponentially decaying terms $\exp\left(-t/\tau_d\right)$ with $\tau_d\propto\frac{\gamma_0}{E_0T}$ remain its distinctive feature.
From that viewpoint, the TLS effectively assumes that the same exponentially decaying terms dominate $C_{jj}(t)$ even in the statically disordered Peierls model with white noise.
This suggests that $\tau_d$ depends linearly on $\gamma_0$ in the strong-damping limit.

\section{Coarse-grained nature of short-time dynamics for the Drude--Lorentz spectral density}
\label{App:coarse-grained-DL}
Here, we explore how $\mathrm{Im}\:\mathcal{C}_\mathrm{DL}(t=0)\neq 0$ yields problems with the equilibrium expectation value
\begin{equation}
\label{Eq:B1}
\left\langle j_\mathrm{e-ph}^2\right\rangle=\sum_{\substack{q_2m_2\\q_1m_1}}\mathrm{Tr}_S\left\{J_{q_2}J_{q_1}\mathrm{Tr}_B\left\{f_2f_1\rho_\mathrm{tot}^\mathrm{eq}\right\}\right\},   
\end{equation}
which should be purely real (and positive).
In Eq.~\eqref{Eq:B1} and in the following, we abbreviate $f_{q_im_i}\equiv f_i$.

Using Eqs.~\eqref{Eq:GWT_gtr} and~\eqref{Eq:def_equal_time_gtr}, we evaluate $\mathrm{Tr}_B\left\{f_2f_1\rho_\mathrm{tot}^\mathrm{eq}\right\}$ as
\begin{equation}
\label{Eq:B2}
\begin{split}
    &\mathrm{Tr}_B\left\{f_2f_1\rho_\mathrm{tot}^\mathrm{eq}\right\}=\mathrm{Tr}_B\left\{F_{\mathbf{0}_2^+}^{(1)}f_1\rho_\mathrm{tot}^\mathrm{eq}\right\}\\&=\rho_{\mathbf{0}_{21}^+}^{(2,\mathrm{eq})}+\delta_{m_1\overline{m_2}}\eta_{q_2q_1m_2}\rho_\mathbf{0}^{(0,\mathrm{eq})}.
\end{split}
\end{equation}
Instead of Eq.~\eqref{Eq:GWT_gtr}, we could have used the generalized Wick's theorem in the form~\cite{MolPhys.116.780,JChemPhys.157.170901}
\begin{equation}
\begin{split}
    &\mathrm{Tr}_B\left\{f_{qm}F_\mathbf{n}^{(n)}\rho_\mathrm{tot}^\mathrm{eq}\right\}=\\
    &\rho_{\mathbf{n}_{qm}^+}^{(n+1,\mathrm{eq})}+\sum_{q'm'}n_{q'm'}\langle f_{qm}f_{q'm'}\rangle_B^<\rho_{\mathbf{n}_{q'm'}^-}^{(n-1,\mathrm{eq})},
\end{split}
\end{equation}
with [see Eq.~\eqref{Eq:dissipaton_CF_bwd}]
\begin{equation}
    \langle f_{qm}f_{q'm'}\rangle_B^<=\left\langle f_{qm}^{(I)}(0)f_{q'm'}^{(I)}(0_+)\right\rangle_B=\delta_{m\overline{m'}}\eta_{\overline{q'}\:\overline{q}\:\overline{m'}}^*,
\end{equation}
to obtain
\begin{equation}
\label{Eq:B3}
\begin{split}
    &\mathrm{Tr}_B\left\{f_2f_1\rho_\mathrm{tot}^\mathrm{eq}\right\}=\mathrm{Tr}_B\left\{f_2F_{\mathbf{0}_1^+}^{(1)}\rho_\mathrm{tot}^\mathrm{eq}\right\}\\&=\rho_{\mathbf{0}_{21}^+}^{(2,\mathrm{eq})}+\delta_{m_1\overline{m_2}}\eta_{\overline{q_2}\:\overline{q_1}\:\overline{m_2}}^*\rho_\mathbf{0}^{(0,\mathrm{eq})}.
\end{split}
\end{equation}
Inserting Eqs.~\eqref{Eq:B2} or~\eqref{Eq:B3} into Eq.~\eqref{Eq:B1}, and requiring that the two results for $\langle j_\mathrm{e-ph}^2\rangle$ be identical, we obtain
\begin{equation}
\label{Eq:B4}
\sum_{q_2q_1m_2}\left(\eta_{q_2q_1m_2}-\eta_{\overline{q_2}\:\overline{q_1}\:\overline{m_2}}^*\right)\mathrm{Tr}_S\left\{J_{q_2}J_{q_1}\rho_\mathbf{0}^{(0,\mathrm{eq})}\right\}=0.
\end{equation}
Recalling that $\eta_{q_2q_1m_2}=\frac{\delta_{q_1\overline{q_2}}}{N}c_{m_2}$, we transform Eq.~\eqref{Eq:B4} into
\begin{equation}
\label{Eq:B5}
\sum_m\left(c_m-c_m^*\right)\sum_q\mathrm{Tr}_S\left\{J_q^\dagger J_q\rho_\mathbf{0}^{(0,\mathrm{eq})}\right\}=0.
\end{equation}
To obtain the first factor of Eq.~\eqref{Eq:B5}, we perform the change of the dummy index $\overline{m}\leftrightarrow m$, which is permissible even for finite $K$ because the truncated DEOM always includes pairs of complex-conjugated components.
Each $q$-dependent term $\mathrm{Tr}_S\left\{J_q^\dagger J_q\rho_\mathbf{0}^{(0,\mathrm{eq})}\right\}$ is positive because $\rho_\mathbf{0}^{(0,\mathrm{eq})}$ is the equilibrium reduced density matrix, while the operator $J_q^\dagger J_q$ is positive.
We thus conclude that the two ways of evaluating $\langle j_\mathrm{e-ph}^2\rangle$ yield the same result under the condition that
\begin{equation}
\label{Eq:B6}
    \sum_m\mathrm{Im}\:c_m=0.
\end{equation}
While Eq.~\eqref{Eq:B6} is satisfied for the BO SD, see Appendix~\ref{App:c_m_mu_m}, for the DL SD we have
\begin{equation}
    \sum_m\mathrm{Im}\:c_m^\mathrm{DL}=-E_0\gamma_\mathrm{DL}\neq 0.
\end{equation}
This inconsistency of the DL model is resolved using Ishizaki's procedure.~\cite{JPhysSocJpn.89.015001}

\newpage

\bibliography{aipsamp}

\end{document}